\documentclass[sigconf,10pt]{acmart}
\renewcommand\footnotetextcopyrightpermission[1]{} 

\usepackage{amsfonts,balance,bbold,bm,color,enumitem,hyperref,mathtools,microtype,multirow,natbib,soul,subcaption,tabularx,xcolor,xspace}
\usepackage[linesnumbered,ruled,vlined]{algorithm2e}
\usepackage{pgfplots}
\pgfplotsset{compat=1.18}

\newcommand{\sn}{\textbf{\textit{MotionLeaf}}\xspace}

\begin{document}

\title[\sn]{\sn:  Fine-grained Multi-Leaf Damped Vibration Monitoring for Plant Water Stress using Low-Cost mmWave Sensors}
\date{April 2024}

\author{Mark Cardamis}
\email{m.cardamis@unsw.edu.au}
\orcid{0000-0001-7896-5038}
\affiliation{%
  \institution{University of New South Wales}
  \streetaddress{K17, Kensington Campus}
  \city{Sydney}
  \state{NSW}
  \country{Australia}
  \postcode{2052}
}

\author{Chun Tung Chou}
\email{c.t.chou@unsw.edu.au}
\orcid{0000-0003-4512-7155}
\affiliation{%
  \institution{University of New South Wales}
  \city{Sydney}
  \state{NSW}
  \country{Australia}
}

\author{Wen Hu}
\email{wen.hu@unsw.edu.au}
\orcid{0000-0002-4076-1811}
\affiliation{%
  \institution{University of New South Wales}
  \city{Sydney}
  \state{NSW}
  \country{Australia}
}

\renewcommand{\shortauthors}{Cardamis et al.}

\begin{abstract}
In this paper, we introduce \sn, a novel mmWave-base multi-point vibration frequency measurement system that can estimate plant stress by analyzing the
the surface vibrations of multiple leaves. \sn features a novel signal processing pipeline that accurately estimates fine-grained damped vibration frequencies based on noisy micro-displacement measurements from a
mmWave radar. Specifically we explore the Interquartile Mean (IQM) of coherent phase differences from neighboring Frequency-Modulated Continuous Wave (FMCW) radar chirps to calculate micro-displacements. Furthermore,  
we use the measurements from multiple received antennas in the radar to
estimate the vibration signals of different leaves via a Blind Source Separation (BSS) method. Experimental results demonstrate that \sn can accurately measure the frequency of multiple leaves in a plant with average error of 0.0176 Hz, which is less than 50\% of that (0.0416 Hz) of the state-of-the-art approach (mmVib~\cite{jiang_mmvib_2020}). Additionally, the estimated natural vibration frequencies from \sn are shown to be an excellent feature
to detect the water stress in the plant during 7-day drought experiments.
\end{abstract}



\maketitle

\section{Introduction}

Water is essential for the survival of plants, and the water content in plant leaves is a crucial indicator of plant health and stress. Water stress can have a significant impact on crop yields and can lead to reduced growth and productivity. In order to effectively manage water stress, it is important to have accurate and non-destructive methods for measuring water content in plants.

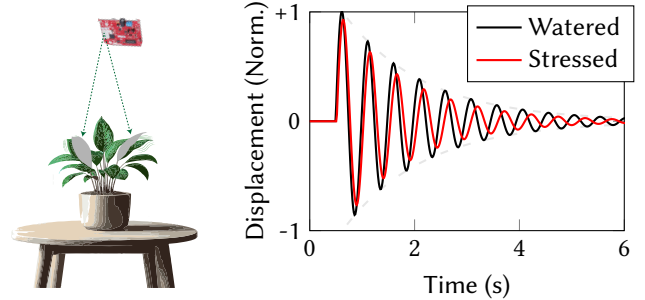
\begin{figure}[t]
    \centering
    \subfloat[\sn in operation\label{fig:introradarsignalmultipleleavesplant}]
    {\includegraphics[trim={0cm 0.2cm 0cm 0cm},clip,height=0.47\linewidth]{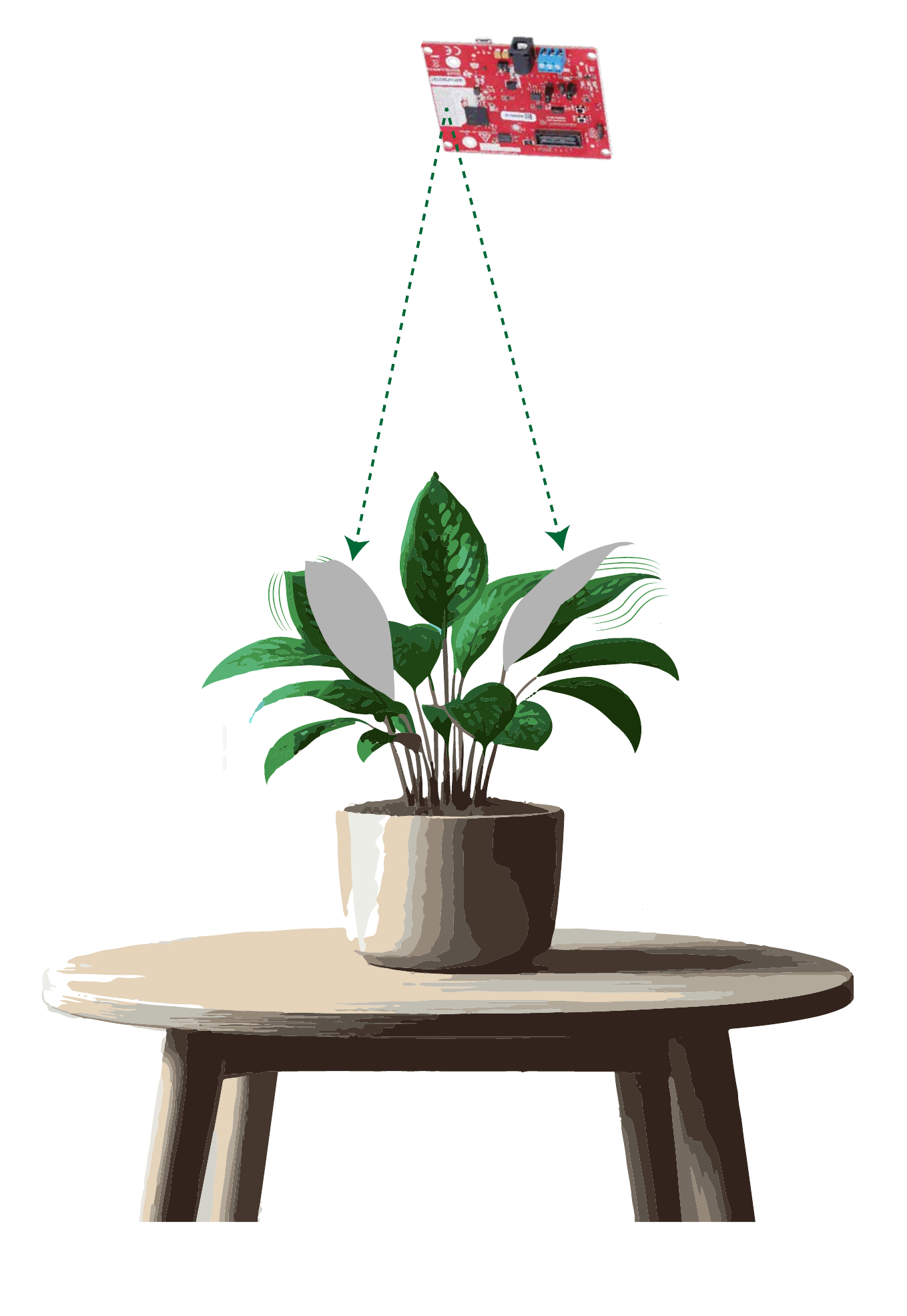}}
    \hspace*{0.5em}
    \subfloat[Damped leaf vibration frequency patterns. Frequency: Watered/healthy (2.03 Hz), Stressed (1.95 Hz)\label{fig:methodstemdisplacementtime}]{
        \begin{tikzpicture}[every node/.style={font=\sffamily}]
          \begin{axis}[
            width=0.68*\linewidth, height=0.53*\linewidth,
            xlabel={Time (s)},
            ylabel={Displacement (Norm.)},
            ylabel style={yshift=-0.5em},
            xmin=0, xmax=6,
            ymin=-10, ymax=10,
            ytick={-10,0,10}, 
            yticklabels={-1,0,+1},
            legend pos=outer north east,
            legend cell align={left},
            legend style={
                font=\sffamily\small,
                at={(0.5,1.03)},
                anchor=north west
            },
            legend entries={Watered, Stressed},
            every axis plot/.append style={thick},
            ]
            \addplot[black, domain=0:14, samples=1000] 
                {(x <= 0.5) * 0.01 + (x > 0.5) *15 * exp(-0.65*x) * cos(deg(4.06*pi*x - pi/2-0.01))};
            \addplot[red, domain=0:14, samples=1000] 
                {(x <= 0.51) * 0 + (x > 0.51) *15 * exp(-0.75*x) * cos(deg(3.9*pi*x - pi/2 - 0.01))};
            \addplot[black, loosely dashed, domain=0:14, samples=500, opacity=0.1] 
                {15 * exp(-0.65*x)};
            \addplot[black, loosely dashed, domain=0:14, samples=500, opacity=0.1] 
                {-15 * exp(-0.65*x)};
          \end{axis}
        \end{tikzpicture}}
    \caption{Multi-leaf vibration frequency monitoring with \sn. }
    \label{fig:teaser}
    \Description{}
    \vspace{-1em} 
\end{figure}
Figure~\ref{fig:teaser} illustrates an example set up of \sn (a) and the vibration frequency changes between a well-watered and a water-stressed plant (b). The graph shows the displacement over time, with the well-watered plant exhibiting a higher vibration frequency (2.03Hz) compared to the water-stressed plant (1.95Hz). This difference in vibration behavior highlights the impact of water stress on the mechanical properties of the plant.

Plants naturally vibrate due to environmental factors such as wind, rain, and temperature changes. The vibration properties of plants can be influenced by their physiological state, including water status, nutrient availability, and disease resistance. Previous studies \cite{sano_estimation_2015, de_langre_nondestructive_2019,caicedo-lopez_effects_2020} have shown that analyzing the vibration properties of plants can provide information about their health status. However, these studies are limited to small sample sizes in controlled experimental conditions. 
Laser-based methods \cite{sano_estimation_2015, caicedo-lopez_effects_2020} offer high accuracy but are costly and require smooth, flat surfaces. Rough, uneven leaf surfaces can reduce accuracy due to height variations as the leaf moves during plant motion. Vision-based solutions are more cost effective \cite{de_langre_nondestructive_2019}, however they are sensitive to environmental illumination conditions and generally have poor performance at night-time. Accelerometers need in-situ installation so are typically only used on trunks and larger branches to track coarse seasonal changes \cite{gougherty_estimating_2018}. 
Radar systems typically have a larger field of view compared to lasers or cameras, allowing for the simultaneous monitoring of multiple plants which is advantageous for large-scale monitoring. Radar can also work in the dark, enabling 24-hour monitoring of plant vibrations and thus the study of diurnal patterns.

Turgor pressure significantly influences the mechanical properties of plant tissues, affecting stiffness and elasticity \cite{caliaro_effect_2013}. During water stress, a plant experiences water loss from its cells, resulting in reduced cell turgor pressure \cite{gonzalez-rodriguez_turgidity-dependent_2016}. This decrease in turgor pressure leads to structural changes in the plant’s tissues. Increased turgor pressure causes the cell wall to expand, leading to stiffer tissue, whereas decreased turgor pressure results in the contraction of the cell wall, reducing tissue stiffness.
The natural frequency represents the plant’s inherent vibration rate. When turgor pressure rises, the natural frequency tends to increase. Conversely, reduced turgor pressure leads to a decrease in the natural frequency.

Leveraging plant vibrations for turgor pressure detection offers several advantages over conventional methods. Firstly, it’s non-invasive, eliminating the need for physical contact or destructive sampling. Secondly, it’s sensitive to subtle shifts in turgor pressure, unlike visual inspection or thermal imaging. Lastly, it enables real-time, continuous plant health monitoring. Although the mmWave-based wireless vibration sensing has made tremendous progress over the last few years~\cite{jiang_mmvib_2020}, it encounters several unique challenges in plant vibration sensing as follows:
\begin{itemize}
    \item Multiple-point vibration information acquisition: A plant normally has multiple leaves, each of which may vibrate in a different frequencies and amplitudes (micro-displacement). Therefore, a mmWave radar may receive the wireless signal affected by multiple vibration sources. Hence, the primary goal of \sn is designed to collect vibration information from multiple points. 
    \item Micro displacement measurement: the damped vibration of a leaf is weak and its duration is short (see Figure~\ref{fig:methodstemdisplacementtime} for an example). Therefore, it susceptible to external disturbances of uncertain noise. However, the frequency difference between stressed and non-stressed plant is subtle (e.g., less than 0.1Hz, see Figure~\ref{fig:methodstemdisplacementtime} for an example). Hence, it is a challenging technical task to measure the vibration frequencies accurately from the leaf.
    
\end{itemize}

In this study, we introduce \sn, an innovative non-contact system that leverages the natural vibration patterns of leaves to detect turgor pressure and infer plant health. Our system explores the properties of the reflected signals in the In-phase and Quadrature ($I/Q$) domain to exploit the inherent consistency among these signals, accurately recovering the vibration characteristics of multiple leaves. Our contributions can be summarized as follows:
\begin{itemize}
  \item We propose a mmWave-based plant stress detection system, called \sn, which analyzes the reflected signal from multiple vibrating leaves to determine essential plant dynamics. \sn is a non-invasive method that eliminates the need for physical contact with the plant, increasing the scalability of the solution. 
  \item \sn features a novel signal processing pipeline which effectively removes noise, static objects, and other unwanted signals to recover individual vibration sources.
  Specifically, we introduce a BSS algorithm to separate the vibration signals of different leaves from the wireless signal received at the multiple antenna of a radar.
  Furthermore, we explore the coherent phase difference from the neighboring chirps to calculate the micro-displacements of weak leaf vibration signals.
  \item We implement \sn using a commercial off-the-shelf (COTS) mmWave radar and evaluate its performance in a real-world protected cropping farm. The results show that \sn accurately measures leaf vibration with a Mean Absolute Error (MAE) error of 0.0176 Hz , which is less than 50\% of that (0.0416 Hz) of state-of-the-art approach mmVib~\cite{jiang_mmvib_2020}.
\end{itemize}


\section{Background and Preliminaries}\label{sec:background}

\subsection{Mechanical vibration systems}

\begin{figure}[htbp]
  \centering
  \subfloat[Spring-mass-damper\label{fig:backgroundspringmassdamper}] {\includegraphics[trim={0cm 0cm 0cm -2cm},clip,height=0.21\linewidth]{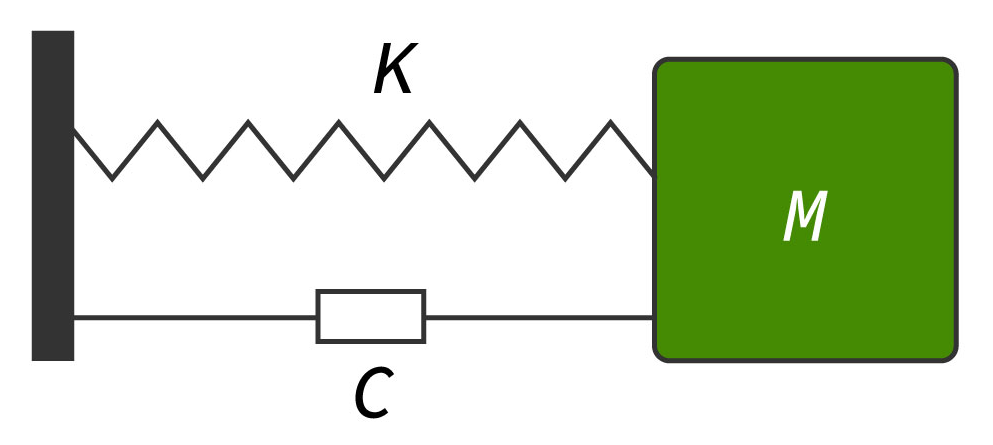}}
  \hspace*{1em}
  \subfloat[Leaf-stem cantilever beam
  \label{fig:backgroundplantleafcantileverbeam}]{\includegraphics[trim={0cm 1cm 0cm 0cm},clip,height=0.25\linewidth]{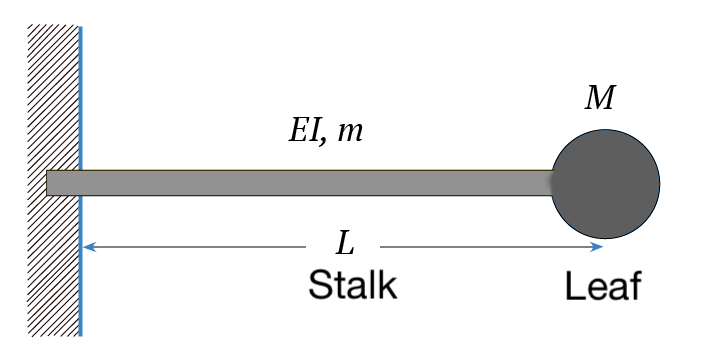}}
  \caption{Simple mechanical model of a plant structure}
  \label{fig:backgroundplantleafplantstructure}
  \Description{}
\end{figure}

Plants behave like damped harmonic oscillators~\cite{spatz_oscillation_2013, de_langre_effects_2008} which can be modeled using mechanical theory, shown in Figure~\ref{fig:backgroundspringmassdamper}, with an example response shown in Figure~\ref{fig:methodstemdisplacementtime}. A plant stem or branch can be thought in terms of a cantilever beam, where one end is fixed and the other is free to move in space~\cite{de_langre_effects_2008}, as shown in Figure~\ref{fig:backgroundplantleafcantileverbeam}. When there is a weight, such as a fruit or leaf, at the tip of the plant structure, it introduces additional forces and moments. This cantilever beam structure will have a mode of vibration defined by its oscillation (natural frequency) and how it decays in time (damping ratio), example model shown in Figure~\ref{fig:backgroundspringmassdamper}. The mode of vibration refers to a particular shape of free motion that oscillates in time before dying out. Modes are fundamental to linear vibrations as they are used to reconstruct the motion of a system and can be calculated using the second-order linear ordinary differential equation, shown in Equation~\ref{equ:vibrationmode} below:

\begin{equation}\label{equ:vibrationmode}
    m\ddot{X} + c\dot{X} + kX = F,
\end{equation}
where $m$ is a mass, $X(t)$ is the modal motion of interest, $\dot{X}$ and $\ddot{X}$ are the corresponding velocity and acceleration respectively, $c$ is the damping coefficient, $k$ is the stiffness, and F(t) is the applied load. 
If the system is static, Equation~\ref{equ:vibrationmode} can be simplified to: 
\begin{equation}\label{equ:vibrationmodestatic}
    kX = F.
\end{equation}

In our simple one-dimensional cantilever beam, shown in Figure~\ref{fig:backgroundplantleafcantileverbeam}, the spring constant $k$ [N/m] can be calculated~\cite{denny_biology_1988} using Equation~\ref{equ:springconstant} below: 

\begin{equation}\label{equ:springconstant}
    k = \frac{3 E I}{L^3},
\end{equation}
where $E$ [Pa] represents the Young's modulus,
$I$ [$m^4$] represents the secondary moment of the cross-sectional area, and $L$ [m] represents the length of the system.

When an impulse is applied to the system, such as a hammer input or a pull and release technique~\cite{spatz_oscillation_2013}, the natural frequency of the mode can be calculated using Equation~\ref{equ:naturalfrequency}. 

\begin{equation}\label{equ:naturalfrequency}
    \omega_n = \sqrt{\frac{k}{m}} = \sqrt{\frac{3 E I}{L^3 m}},
\end{equation}
where $\omega_n$ [rad s$^{-1}$] represents the natural frequency. 

The next section explores plant dynamics and examines how various biological systems affect the vibration modes.

\subsection{Plant dynamics}
\subsubsection{The relationship of turgor pressure on plant dynamics}\label{sec:turgorpressure}

Turgor Pressure is the force exerted by stored water against a cell wall within a plant. It is a key component of plant physiology that maintains the structural integrity of the plant. High Turgor Pressure causes the cell walls to expand and provides a rigid structure to the plant where as Low Turgor Pressure causes the cells walls to shrink and results in a wilted plant.

The elastic modulus is a measure of a material's resistance to deformation and is directly influenced by turgor pressure. Nilsson \cite{nilsson_relation_1958} derived a relationship between the elastic modulus ($E$) and turgor pressure ($\Psi_p$), with $E$ being a linear function of $\Psi_p$. This relationship links the plant's biological property ($\Psi_p$) to the plant's mechanical property $E$, with the tissue pressure directly effecting the plant stiffness.

These changes in stiffness influence the plant's natural frequency of vibration, calculated using Equation~\ref{equ:naturalfrequency}. The natural frequency is the rate at which a plant vibrates when subjected to an external force. An increase in turgor pressure typically leads to an increase in the natural frequency due to the higher stiffness, whereas a decrease in turgor pressure results in a lower natural frequency, as the plant becomes more flexible and less resistant to external forces.

Understanding the relationship between turgor pressure and plant dynamics is critical for developing non-invasive methods to monitor plant health. In the following section we discuss how to measure the mechanical beam properties of a plant leaf and illustrate the inverse relationship between plant stiffness and water stress.

\subsubsection{Leaf-stem beam parameters }\label{subsubsec:methodologyleafstiffness}
\begin{figure}[htbp]
  \centering
  \subfloat[Measure spring constant
    \label{fig:methodspringconstantleafsetup}]{\includegraphics[trim={3cm -1cm 2cm 0cm},clip,height=0.34\linewidth]{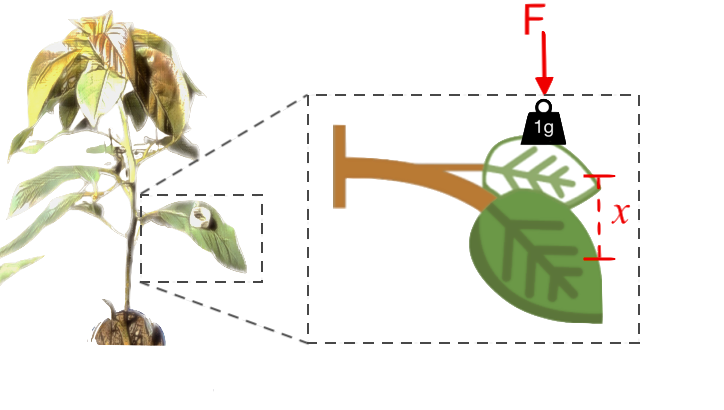}}
  \hspace*{0.1em}
  \subfloat[Spring constant change \label{fig:methodspringconstantleaftable}]{
    \begin{tikzpicture}[every node/.style={font=\sffamily}]
      \begin{axis}[
          width=0.55\linewidth, height=0.4\linewidth,
          xlabel={Days without water},
          xlabel style={xshift=-0.5em,yshift=0.2em},
          ylabel={$k$ (N/m)},
          ylabel style={yshift=-0.5em},
          ymin=0.8, ymax=1.8,
          xmin=0, xmax=12,
          scatter/classes={
              a={mark=*,blue}
          },
          scatter,
        ]
        \addplot[scatter, only marks, scatter src=explicit symbolic]
        coordinates {
          (0,1.7)[a]
          (1,1.6)[a]
          (2,1.6)[a]
          (6,1.3)[a]
          (8,1.2)[a]
          (9,1.1)[a]
          (10,1.0)[a]
          (11,1.0)[a]
        };
      \end{axis}
  \end{tikzpicture}}
    \vspace{-1em}
  \caption{Leaf spring constant test experiment}
  \label{fig:methodspringconstantleaf}
  \Description{}

\end{figure}
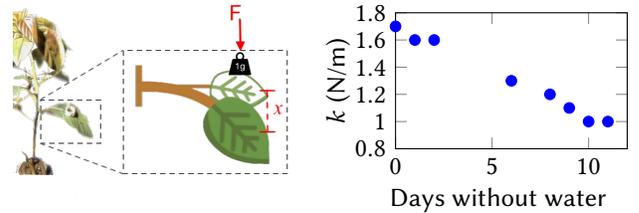

As previously discussed in Section~\ref{sec:turgorpressure}, the plant stiffness is influenced by turgor pressure, which decreases when the plant experiences water stress. Measuring the kinematics of the leaf-stem can be done through a sample tests~\cite{doare_effect_2004}. Under a static load, using Equation~\ref{equ:vibrationmodestatic}, the spring constant $k$ can be determined by measuring the displacement of the beam against a input force. We conducted a water stress study by placing a 1 g weight on a leaf of a plant (see Figure~\ref{fig:methodspringconstantleafsetup}), and  
measuring its displacement from its resting position after vibration. Figure~\ref{fig:methodspringconstantleaftable} shows the overall trend was as expected from the literature~\cite{caliaro_effect_2013} where the leaf-stem stiffness decreased with drought stress. 

Futhermore, the turgor pressure follows daily patterns aligned with the 24-hour day/night cycle, known as the diurnal cycle ~\cite{ache_stomatal_2010}. This behavior offers valuable insight into the plant's dynamics throughout the day and will be explored further in the next section.

\subsubsection{Diurnal (Day/Night) Cycle}\label{sec:diurnalcycle}

Plants undergo significant physiological changes between day and night. This cycle profoundly impacts various plant functions, including turgor pressure, water content, and mechanical properties, which in turn affect the plant's natural vibration frequencies.

Photosynthesis occurs during the day, leading to water uptake and distribution throughout the plant tissues. At night, transpiration rates decrease, and the plant redistributes and conserves water within its tissues. When the soil water content is insufficient, the plant fails to recover at night, and the frequency difference between day and night becomes smaller, which indicates the plant is under stress~\cite{sano_estimation_2015}.

The difference in diurnal frequency can thus serve as a valuable indicator of plant health, particularly in terms of water stress. By monitoring the frequency differences between day and night, we can assess the plant's ability to recover and maintain its physiological functions. A higher diurnal (day-night) frequency difference suggests a healthy plant with efficient water uptake and redistribution mechanisms, while a minimal frequency difference indicates plant stress.

\subsubsection{Different vibration modes in a plant}\label{subsubsec:vibrationModes}
\begin{figure}[htbp]
  \centering
  \subfloat[Frequency of vibration modes
  \label{fig:backgroundtreevibrationmodes}]{\includegraphics[height=0.32\linewidth]{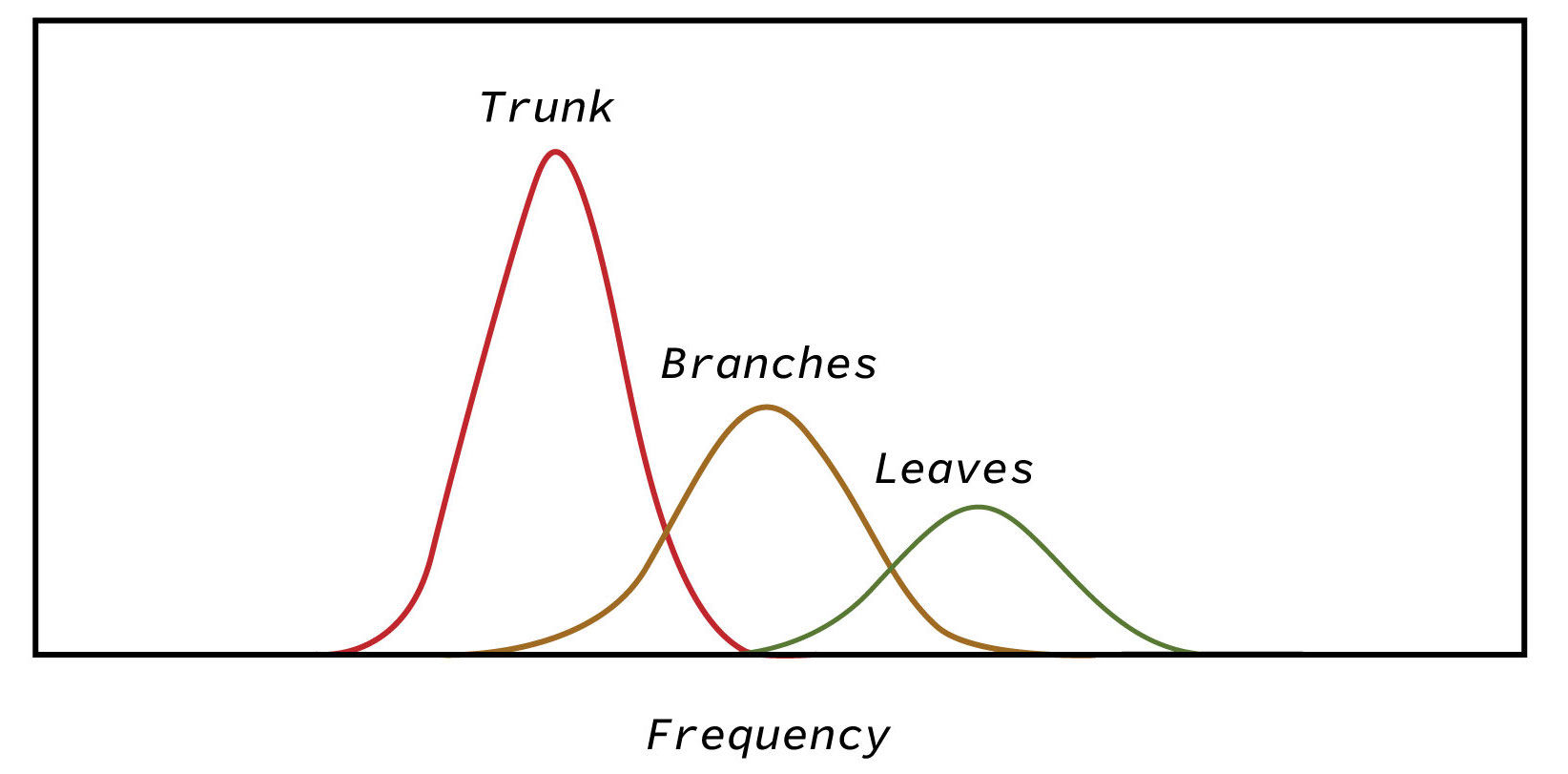}}
  \hspace*{0.5em}
  \subfloat[Hammer input \label{fig:methodvibrationimpulseinput}]{\includegraphics[trim={0cm 1.3cm 0cm 0cm},clip,height=0.33\linewidth]{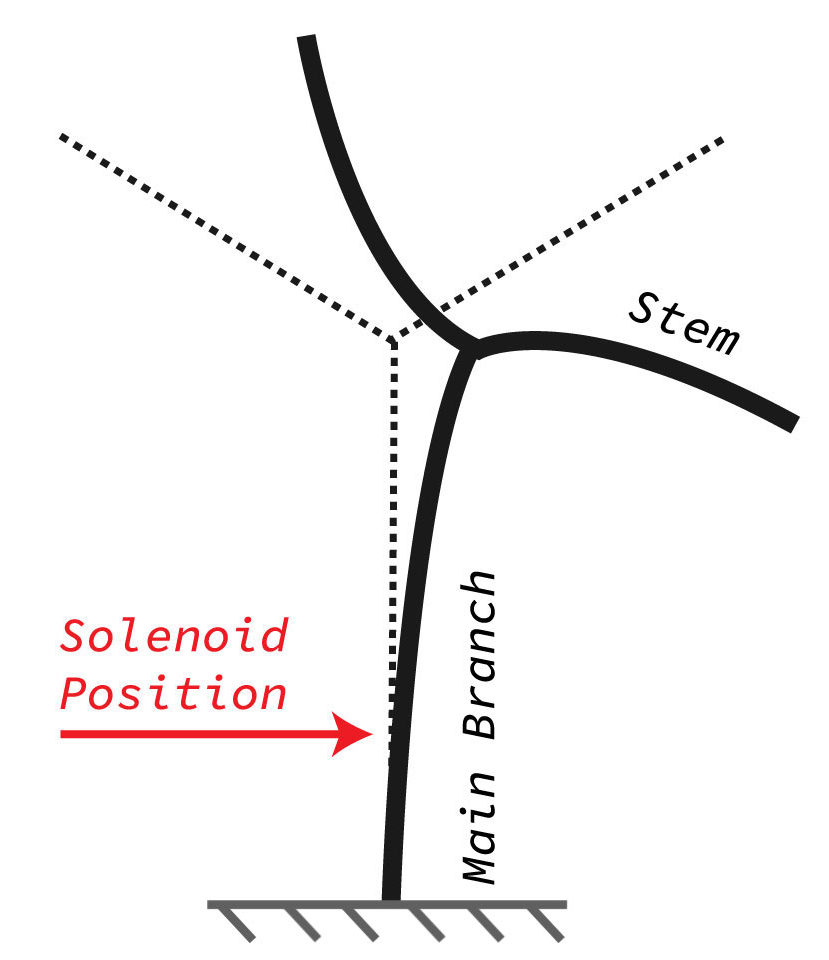}}
  \vspace{-1em}
  \caption{Measuring vibration modes in the plant}
  \label{fig:methodstemdisplacement}
  \Description{}
  \vspace{-1em}
\end{figure}

Plants can exhibit various vibration modes in different parts of their structure, depending on their size, flexibility, and the forces acting upon them. The vibration frequencies across the plant are typically on the order of 1 Hz or 10 Hz \cite{de_langre_plant_2019}. Figure~\ref{fig:backgroundtreevibrationmodes} illustrates the first-order vibration mode in the trunk, the second-order vibration mode in the branches, and the third-order vibration mode in the leaves.

\textit{Trunk vibration:} It corresponds to the lowest resonance frequency in a plant and is effective for detecting seasonal changes, such as spring or autumn, where influences from the leaves can be measured~\cite{gougherty_estimating_2018}.
\textit{Branch vibrations:} Plant branches can exhibit bending and swaying modes in response to external forces like wind or touch. Branches have higher resonant frequencies and represent the second-order vibration modes.
\textit{Leaf vibrations:} Leaves, being the most flexible and lightweight parts of a plant, have the highest resonant frequencies. They represent the third-order vibration mode, as shown in Figure~\ref{fig:backgroundtreevibrationmodes}. Since leaves are influenced by physiological factors like turgor pressure and hydraulic conductance~\cite{gougherty_estimating_2018}, they serve as ideal targets for assessing plant health. 

Furthermore, this behavior implies that an effective plant monitoring method may need to separate different orders of vibration modes to obtain fine-grained vibration information (e.g., frequency). To address this, \sn formulates it as a BSS problem and proposes a novel ICA-based method to unmix aggregated vibration information (see Section~\ref{subsubsec:datapipelinestep3} for details).

\subsubsection{Vibration impulse force}
In order to measure the different vibration modes in the plant, i.e., trunk, branch and leaf stem, a sufficient impulse force should be applied to provide a measurable displacement, example shown in Figure~\ref{fig:methodvibrationimpulseinput}. Too small of an input force will not provide measurable displacement while too high of a force may cause physical mechanical damage~\cite{de_langre_plant_2019}. With the appropriate impulse force, the different vibration modes of the plant can be identified, as shown in Figure~\ref{fig:methodstemdisplacement}. The following section is a primer in using Radar for Plant Vibrations, where we introduce preliminary concepts of the Frequency-Modulated Continuous Wave (FMCW) mmWave signals.

\begin{figure*}[htbp]
  \centering
  \includegraphics[width=\linewidth]{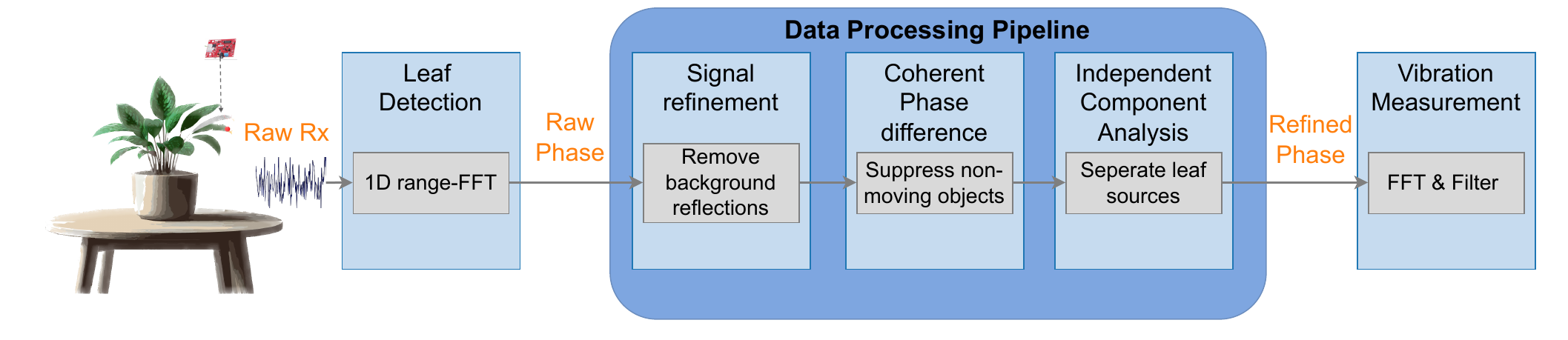}
  \caption{Overview of the \sn workflow}
  \label{fig:methodoverviewsystemworkflow}
  \Description{A block diagram representing the system workflow of \sn}
\end{figure*}

\subsection{Using Radar for Plant Vibrations}

The use of radar-based techniques to measure plant vibrations offers substantial advantages over traditional in-situ sensors:

\begin{itemize}
    \item No Impact on Plant Dynamics: Attached sensors can add mass to the plant altering the vibration properties. Radar-based methods do not affect the plant's physical properties, ensuring that the data collected is a true representation of the plant's natural vibration state. This is particularly important in studies of small or lightweight plants, where the added weight of sensors can significantly alter the plant dynamics~\cite{de_langre_effects_2008}.
    \item Multi-Point Measurements: Unlike traditional sensors that are typically limited to single-point measurements, radar systems have the capability to simultaneously monitor multiple points. Simultaneous multi-leaf tracking can provide a more detailed picture of the plant’s overall condition and detect localized stress' that might be missed with single-point measurements.
\end{itemize}

\subsubsection{FMCW Radar}
FMCW radar works by transmitting a continuous linearly increasing signal called a ``chirp'', whose frequency increases linearly over time. 

These chirps allow the radar to simultaneously measure the distance and velocity of target objects. The basic principle involves transmitting a chirp signal and measuring the delayed ($\Delta t$) reflected signal. The difference between the transmitted and received signal produces a beat frequency which is proportional to the object distance. The beat frequency, $\Delta f$, is calculated as:

\begin{displaymath}
\Delta f = 2 \frac{d_t \cdot BW}{c \cdot T_r}
\end{displaymath}
where $d_t$ is the distance to the target, $BW$ is the bandwidth of the transmitted signal, $c$ is the speed of light, and $T_r$ is the ramp time of the chirp.

Small changes in the position of an object is known as micro-displacements. FMCW radar is capable of detecting micro-displacements by analyzing the phase shift in the received radar signals, discussed further below.

\subsubsection{mmWave Vibration Measurement}
The principle of measuring micro-displacements using mmWave radar is that the phase shifts occur due to slight changes in the signal propagation path.
When a mmWave chirp signal reflects off a vibrating object, the phase of the received signal will vary according to the displacement of the object.

The displacement changes can be observed by tracking the phase value $\varphi$ of the complex $I/Q$ signal captured by the FMCW radar. The relationship between the phase change $\Delta \varphi$ and the displacement $\Delta d$ is shown below:
\begin{displaymath}
\Delta \varphi = 2\pi \cdot \frac{2 \Delta d}{\lambda}
\end{displaymath}
where $\lambda$ is the signal wavelength.

Micro-Doppler techniques have been used to estimate the velocity of an object's movement by performing a Doppler-FFT on the signal. However, when there are multiple objects in the reflected signal, we cannot derive $\Delta d$ from $\Delta \varphi$~\cite{jiang_mmvib_2020}, thus we focus on getting accurate phase value $\varphi$ measurements.

To accurately measure the vibration displacement $\Delta d$, the phase values $\varphi$ must be precisely tracked.  
However, this task is non-trivial due to two main challenges. Firstly, the reflected signal is entangled with multiple vibration sources, not just the target leaf, making it difficult to isolate the relevant signal. Secondly, there is the issue of noise tolerance in the phase values, especially given the short observation window of the damped leaf oscillation, example shown in Figure~\ref{fig:methodstemdisplacementtime}. When the observation window is short, noise can appear as random fluctuations in the phase values, making it difficult to distinguish the true signal from the background noise.

\section{Methodology}\label{sec:methodology}

In this section we introduce \sn, a framework designed to extract the vibration patterns of a multi-leaf plant, with an overview of the system workflow presented in Figure~\ref{fig:methodoverviewsystemworkflow}.
We explore the key modules of the \nameref*{subsec:dataprocessingmethodology}: (1) \nameref*{subsubsec:datapipelinestep0}, (2) \nameref*{subsubsec:datapipelinestep1}, (3) \nameref*{subsubsec:datapipelinestep2} and (4) \nameref*{subsubsec:datapipelinestep3}.

\subsection{Data Processing Pipeline}\label{subsec:dataprocessingmethodology}

\subsubsection{Leaf Detection}\label{subsubsec:datapipelinestep0}
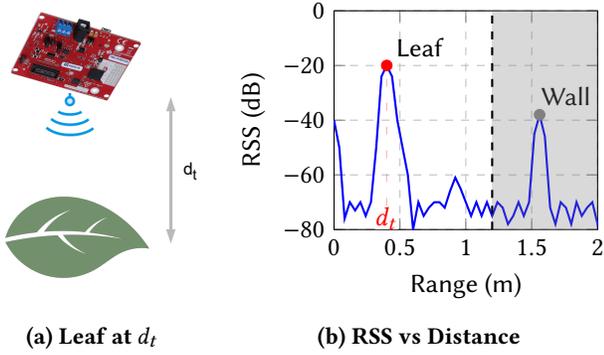
\begin{figure}[htbp]
  \centering
    \subfloat[Leaf at $d_t$ \label{fig:backgroundpowerdistancea}]{\includegraphics[height=0.5\linewidth]{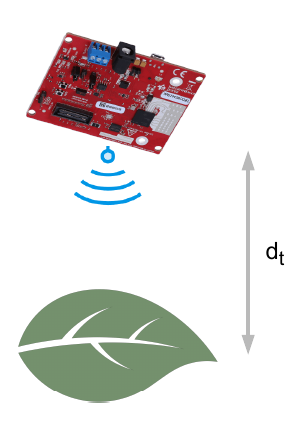}}
  \hspace*{1em}
  \subfloat[RSS vs Distance \label{fig:backgroundpowerdistanceb}]
    {\begin{tikzpicture}[every node/.style={font=\sffamily}]
        \begin{axis}[
            width=0.6*\linewidth, height=0.53*\linewidth,
            xlabel={Range (m)},
            ylabel={RSS (dB)},
            grid=both,
            ymin=-80,
            ymax=0,
            xmin=0,
            xmax=2,
            ymajorgrids=true,
            xmajorgrids=true,
            grid style=dashed,
        ]
        \addplot[blue, thick] coordinates {
            (0.00, -40)
            (0.04, -50)
            (0.08, -76)
            (0.12, -70)
            (0.16, -73)
            (0.20, -70)
            (0.24, -75)
            (0.28, -70)
            (0.32, -50)
            (0.36, -24)
            (0.40, -20)
            (0.44, -24)
            (0.48, -40)
            (0.52, -50)
            (0.56, -60)
            (0.60, -80)
            (0.64, -70)
            (0.68, -75)
            (0.72, -72)
            (0.76, -70)
            (0.80, -70)
            (0.84, -72)
            (0.88, -66)
            (0.92, -61)
            (0.96, -65)
            (1.00, -70)
            (1.04, -75)
            (1.08, -70)
            (1.12, -75)
            (1.16, -70)
            (1.20, -75)
            (1.24, -70)
            (1.28, -72)
            (1.32, -78)
            (1.36, -75)
            (1.40, -70)
            (1.44, -75)
            (1.48, -70)
            (1.52, -45)
            (1.56, -38)
            (1.60, -46)
            (1.64, -72)
            (1.68, -78)
            (1.72, -70)
            (1.76, -72)
            (1.80, -78)
            (1.84, -70)
            (1.88, -75)
            (1.92, -72)
            (1.96, -70)
            (2.00, -78)
            (2.04, -75)
        };
        \addplot[red, only marks, mark=*] coordinates {(0.40, -20)};
        \addplot[gray, only marks, mark=*] coordinates {(1.56, -38)};
        \node at (axis cs:0.40,-20) [anchor=south west] {Leaf};
        \node at (axis cs:1.50,-38) [anchor=south west] {Wall};
        
        \fill[gray, opacity=0.3] (axis cs:1.2, -80) rectangle (axis cs:2, 0);
        \draw[red, loosely dashed, thin, opacity=0.3] (axis cs:0.4, -80) -- (axis cs:0.4, -20);
        \draw[dashed, thick] (axis cs:1.2, -80) -- (axis cs:1.2, 0);
        \node[text=red, fill=white, inner sep=0pt] at (axis cs:0.40,-80) [anchor=south] {$d_t$};

        \end{axis}
    \end{tikzpicture}}
  \caption{mmWave signal reflections}
  \label{fig:backgroundpowerdistance}
  \Description{The graphs show the reflected power (in dBFS) for each range bin and represent the 1D radar output}
  \vspace{-1.0em}
\end{figure}

The radar system continuously transmits chirps and receives reflections from the vibrating object. The received signal is recorded as complex IQ data, where each data point represents a specific time instant. To accurately measure the vibration displacement of the target leaf, we first need to calculate the leaf distance from the radar, as shown in Figure~\ref{fig:backgroundpowerdistancea}. 

A \textit{1D Range-FFT} is performed on the IQ data to convert it from the time domain to the frequency domain, where each discrete frequency bin corresponds to a specific distance (or range bin) from the radar.
Then an \textit{Energy Calculation} is computed using the magnitude squared of the complex FFT output for each range bin, representing the strength of the reflected signal from that range, shown in Figure~\ref{fig:backgroundpowerdistanceb}. 
Finally, the \textit{Range Bin Selection} process identifies the range bin ($d_t$) with the highest energy. Predefined distance limits ensure that the selected bin represents the plant and not other background reflections, such as a wall, as illustrated in Figure~\ref{fig:backgroundpowerdistanceb}. The predefined range limit is based on prior knowledge of the expected location of the target leaf relative to the radar. 

By selecting the range bin with the highest energy, we ensure that the radar system focuses on the strongest reflection, which is most likely from the vibrating leaf. The data from the selected range bin, where the real and imaginary parts correspond to the peak, is passed through to the subsequent \textit{\nameref*{subsubsec:datapipelinestep1}} step.

\subsubsection{Signal Phase Refinement}\label{subsubsec:datapipelinestep1}

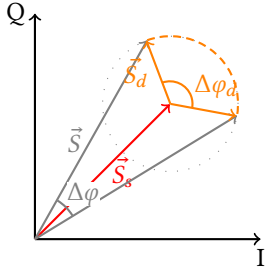
\begin{figure}[htbp]
    \centering
    \begin{tikzpicture}[scale=0.6]
        \draw[thick,->] (0,0) -- (5,0) node[anchor=north] {I};
        \draw[thick,->] (0,0) -- (0,5) node[anchor=east] {Q};
        \draw[gray, loosely dotted] (3,3) circle (1.5);
        \draw[thick,densely dashed,orange] (2.487,4.41) arc[start angle=110,end angle=-10,radius=1.5];
        \draw[thick,->,red] (0,0) -- (2.99,2.99) node[midway, right] {$\vec{S_s}$};
        \draw[thick, ->,gray] (0,0) -- (2.487,4.41) node[midway, left] {$\vec{S}$};
        \draw[thick, ->,gray] (0,0) -- (4.477,2.74);
        \draw[->, thick, orange] (3,3) -- (2.47,4.4) node[midway, left] {$\vec{S_d}$};
        \draw[->, thick, orange] (3,3) -- (4.46,2.73);
        \draw[thick,orange] (2.82,3.47) arc[start angle=110,end angle=-10,radius=0.5] node[midway,  right] {$\Delta \varphi_d$};
        \draw[thick,gray] (0.5,0.855) arc[start angle=58.5,end angle=31.5,radius=1] node[midway,  above right, fill=white, inner sep=0pt] {$\Delta \varphi$};
    \end{tikzpicture}
    \caption{Vibrating object in the $I/Q$ signal domain}
    \label{fig:methodiqsignaldomain}
    \Description{}
\end{figure}

When a radar signal reflects off a target object, there are several components which are mixed together in the received radar signal. The received signal vector $\vec{S}$ comprises of a static vector $\vec{S_s}$ from non-moving objects and a dynamic vector $\vec{S_d}$ from vibrating sources, expressed as $\vec{S}$ = $\vec{S_s}$ + $\vec{S_d}$, as shown in Figure~\ref{fig:methodiqsignaldomain}.

The variations in phase changes $\Delta \varphi_d$ directly correspond to displacement changes by the vibrating object. However, the presence of the static vector component $\vec{S_s}$ reduced the range of the measured phase $\Delta \varphi$ relative to the actual vibrating source phase $\Delta \varphi_d$ as the $I/Q$ signal is not centered at the origin. To accurately measure $\Delta \varphi_d$, we need to remove the background static reflections contained in $\vec{S_s}$.

To retrieve $\Delta \varphi_d$, we utilize a least squares method to fit a circle to the $I/Q$ signal in order to estimate its center~\cite{mikhelson_remote_2011,jiang_mmvib_2020,guo_dancing_2021}. By translating the circle's center to the origin, the static component is effectively removed, thereby isolating the true phase variation corresponding to the vibration displacement. The methodology is summarized as follows:

\begin{itemize}
    \item Modeling Signal Components: Collect $I/Q$ radar data of the vibrating object with static background reflections, where each sample is represented as a point \((I, Q)\) in the complex plane.
    \item Circle Fitting: Apply the least squares method to fit a circle to the $I/Q$ signal using the Levenberg-Marquardt \cite{chernov_least_2005} optimization algorithm and find the circle center $(I_c, Q_c)$.
    \item Center Translation: For each \((I, Q)\) sample, translate the circle's center to the origin to remove the static component from the $I/Q$  signal with each translated sample represented as \((I_i', Q_i')\).
    \item Phase Variation Isolation: The phase variation due to vibration displacement can be isolated by examining the angular changes in these samples. The \((I_i', Q_i')\) samples are converted to phase angles $\varphi_d$ using the arctangent function: 
            \[
            \varphi_d  = \arctan\left(\frac{Q_i'}{I_i'}\right).
            \]       
    \item Vibration Displacement Extraction: Capturing $N$ frames of phase angles $\varphi_d$, where $\{\varphi_{n}\}_{n=1}^{N}$, the change in vibration displacement can be calculated as:
        \begin{equation}
            d_n = \frac{c}{4 \pi f} \text{unwrap}(\varphi_{n}), \quad n \in [1, N].
            \label{eq:d_n}
        \end{equation}
        
\end{itemize}

By following this methodology, the static signals are effectively removed. However, the displacement of a moving leaf often exceeds the wavelength of mmWave radar (e.g., less than 4 mm for a radar operating at 80 GHz), causing discontinuities when the phase exceeds the range $[-\pi, \pi]$. Phase unwrapping is not always effective due to the presence of noise~\cite{itoh_analysis_1982}, especially if the phase values are close to $-\pi$ or $\pi$, which can lead to false vibration peaks during frequency analysis~\cite{xu_simultaneous_2022}. To address this issue, further signal processing techniques are introduced in the following section.

\subsubsection{Coherent Phase Difference}\label{subsubsec:datapipelinestep2}
In this section we discuss the use of a coherent phase difference method to deal with the phase unwrapping issue while also further suppressing the non-moving objects in the radar signal.
In a fast-ramp FMCW radar, a burst of chirps can be sent in short succession to measure an objects velocity. As the transmission time is substantially shorter than the movement time of an object, the coherent phase difference between chirps is relatively unchanged~\cite{hyun_pedestrian_2016}. While this would remove the phase unwrapping problem, as the coherent phase difference between successive chirps would not exceed the $[-\pi, \pi]$ range, the windowing function in the 2D FFT processing can mask weak and slow moving targets~\cite{hyun_pedestrian_2016}.

To obtain a coherent phase measurement from multiple chirps, we analyze the phase difference between consecutive chirps along the number of frames $N$ and apply statistical processing to reduce noise and outliers as follows.

\begin{figure}[htbp]
  \centering
  \subfloat[Unwrap $\varphi_{(l)}$ along Frames
  \label{fig:methodphaseunwrapped}]{\includegraphics[height=0.37\linewidth]{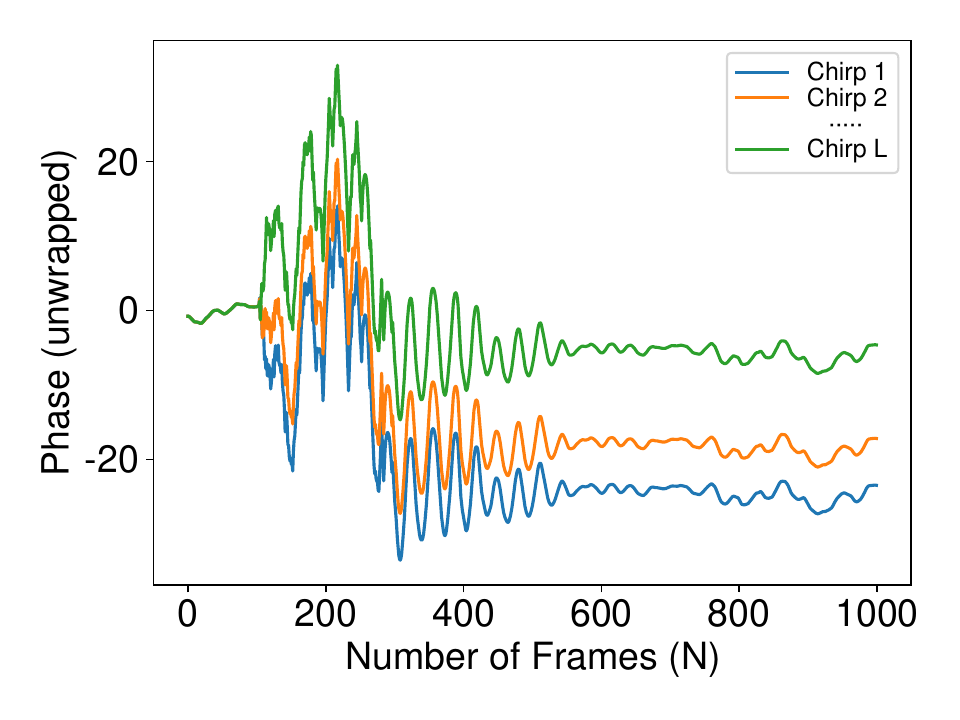}}
  \hspace*{0.2em}
  \subfloat[Diff between $\varphi_{(l)}$ chirps
  \label{fig:methoddeltaphaseunwrapped}]{\includegraphics[height=0.37\linewidth]{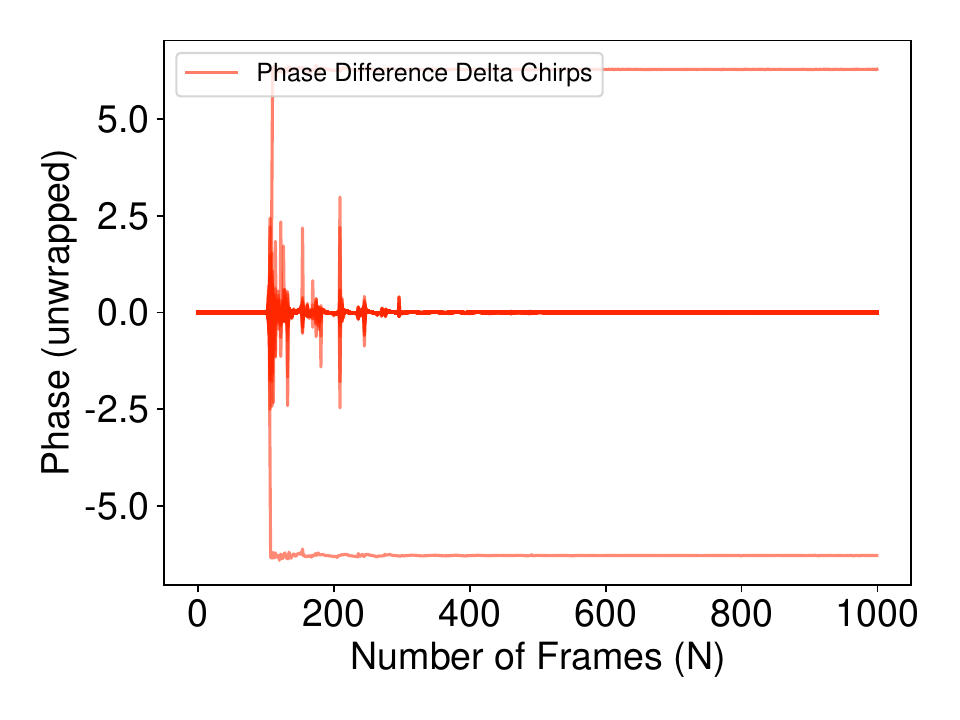}}
  \caption{Phase difference calculation using multiple chirps across a number of frames for FMCW radar.}
  \label{fig:methodphasedifferencecalculation}
  \Description{}
\end{figure}

 
\textbf{Phase Difference Calculation}: In Figure~\ref{fig:methodphasedifferencecalculation}, we illustrate the process of calculating phase differences of mmWave radar measurements between consecutive chirps
from an example leaf damped vibration. For the $l$-th chirp of the phase measurement sequence, where $\varphi_{(l)} =\{\varphi_{l,n}\}_{n=1}^{N}$, we compute the phase difference $\Delta \varphi_{(l)}$ between consecutive chirps as:
    \[
    \Delta \varphi_{(l)} = \text{unwrap}(\varphi_{l+1}) - \text{unwrap}(\varphi_l), \quad l \in [1, L-1]
    \]
where $\varphi_{(l)}$ is the phase measurement sequence for the $l$-th chirp.
Figure~\ref{fig:methodphaseunwrapped} displays the unwrapped phase values \(\varphi_{(l)}\) for each chirp \(l\) in the sequence. Each line represents the same leaf vibration, providing a visual representation of the impact of phase noise on the unwrapping process. Figure~\ref{fig:methoddeltaphaseunwrapped} shows the computed phase difference \(\Delta \varphi_{(l)}\) between consecutive chirps. The phase difference highlights the variations in phase caused by the displacement changes, while making it easier to detect phase sequence outliers coming from the unwrapping process.
 
\begin{figure}[htbp]
  \centering
  \subfloat[Histogram of $\Delta \varphi$
  \label{fig:methodhistogramphasedifference}]{\includegraphics[trim={0cm 0cm 0cm 0cm},clip,height=0.37\linewidth]{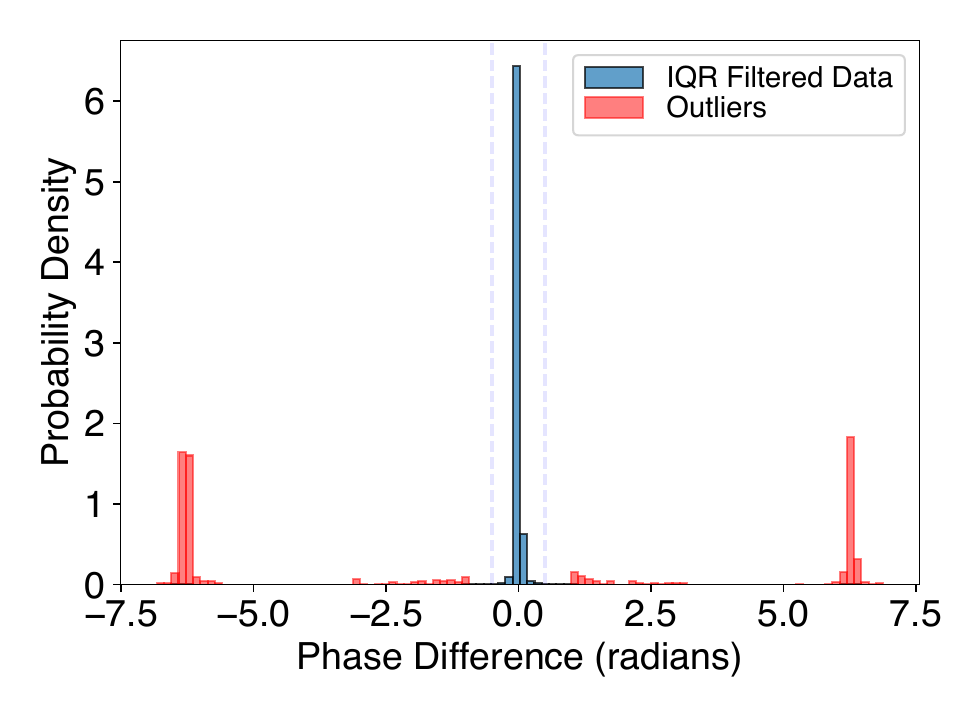}}
  \hspace*{0.2em}
  \subfloat[IQM output of $\Delta \varphi$ \label{fig:methodcoherencephasedifference}]{\includegraphics[height=0.37\linewidth]{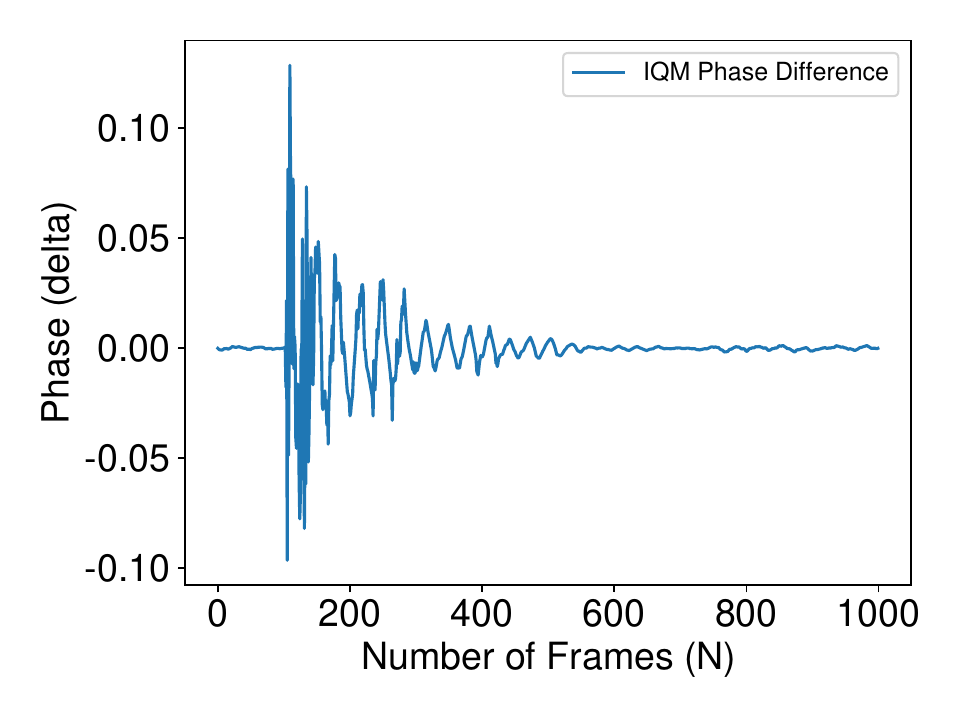}}
  \caption{Outlier detection for the Phase difference data $\Delta \varphi$ and Coherent phase output using IQM.}\label{fig:methodcoherentphasedifferenceprocess}
  \Description{Histogram with Filtered Data and Outliers for the IQM phase difference using chirps axis.}
  \vspace{-1em}
\end{figure}

\textbf{Outlier Reduction Using Interquartile Mean (IQM)}: 
The IQM is a robust statistical method to handle outliers, particularly in data sets with heavy tails or extreme values, as shown in the histogram (see Figure~\ref{fig:methodhistogramphasedifference}). This histogram provides a visual representation of the distribution of phase differences $\Delta \varphi$, where most values are distributed around a central value, and the tails representing outliers.
Specifically, 
IQM has the following features: 

\paragraph{Effective Outlier Detection} IQR focuses on the middle 50\% of the data and excludes both the lower 25\% and the upper 25\%. This is particularly useful for the phase discontinuities present in $\Delta \varphi_{(l)}$.
\paragraph{Improved Data Reliability} By calculating the mean of the values within the IQR, IQM reduces the influence of noisy signals, which leads to a more consistent and reliable phase measurement.
As most of the data points are clustered around zero, by taking the mean of the data within IQR, the IQM method is not influenced by extreme values or outliers in the dataset.
\paragraph{Focus on Central Data} In the histogram, most data points are clustered around zero. The IQM method effectively identifies and excludes outliers on the tail of the distribution. This provides more data points around the zero central peak, which represents the true phase difference of radar signal.

\begin{figure}[htbp]
  \centering
  \subfloat[FFT unwrapped signal.
    \label{fig:methodfftphaseunwrapped}]
    {
      \begin{tikzpicture}
        \node[anchor=south west,inner sep=0] (image) at (0,0) {\includegraphics[height=0.36\linewidth]{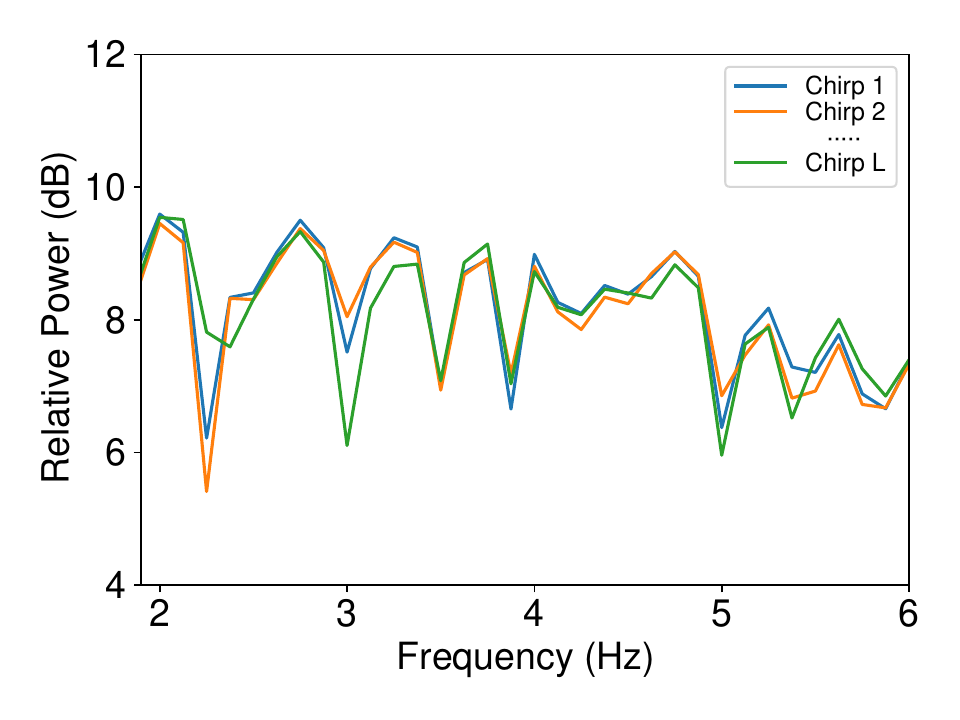}};
        \begin{scope}[x={(image.south east)},y={(image.north west)}]
            \fill[red] (0.558,0.64) circle (2pt); 
        \end{scope}
      \end{tikzpicture}
    }
  \subfloat[FFT IQM signal.
  \label{fig:methodfftcoherencephasedifference}]{\includegraphics[height=0.36\linewidth]{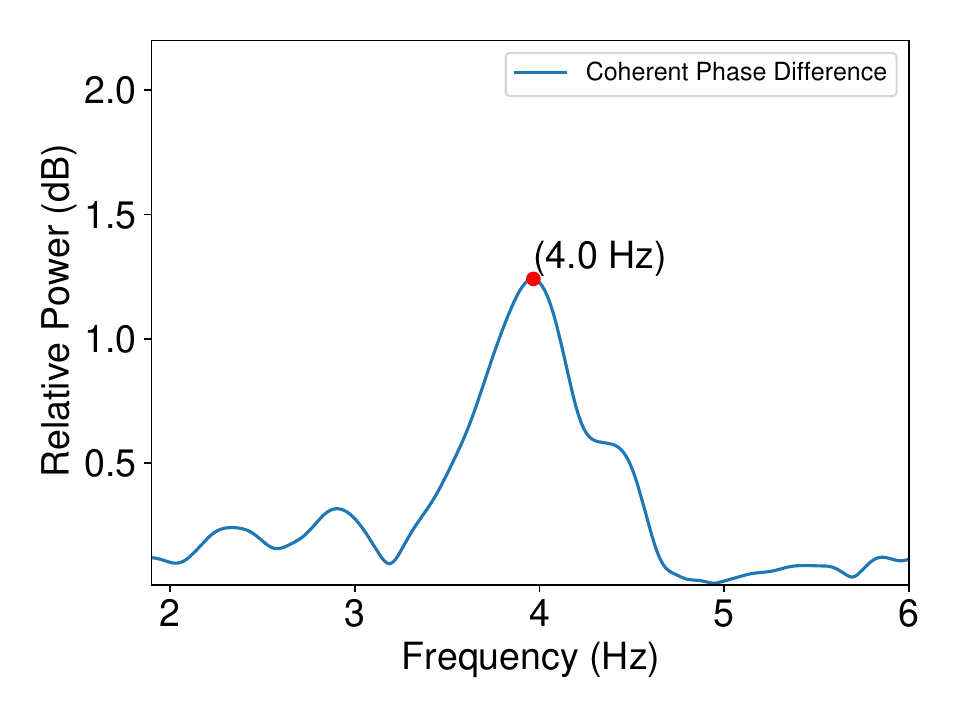}}
  \caption{Frequency analysis of leaf displacement using unwrapped phase signal (Figure~\ref{fig:methodphaseunwrapped}) and IQM phase signal (Figure~\ref{fig:methodcoherencephasedifference})}
  \Description{}
  \label{fig:methodfftphasecomparison}
  \vspace{-1em}
\end{figure}

\textbf{Coherent Phase Output:} Using the IQM-corrected phase differences, one single coherent phase sequence is outputted along the Frame axis. Figure~\ref{fig:methodfftphasecomparison} shows the frequency of an example leaf vibration is more identifiable in the \nameref*{subsubsec:datapipelinestep2} figure (right), compared to the unwrapped phase figure (left).

After collecting the phase differences $\Delta \varphi_{(l)}$ along the Frame axis, we calculate the Interquartile Range (IQR) of the phase differences to identify and exclude outliers. The IQR is the range between the first quartile (Q1) and the third quartile (Q3).
Finally, we compute the IQM by averaging the phase differences that fall within the IQR along the chirp axis, thus reducing the influence of extreme values and noise.

By following this methodology, the phase differences between chirps are effectively processed to produce a coherent phase measurement, minimizing the impact of noise and outliers. This approach ensures more accurate and reliable phase measurements along the slow-time Frame axis. \nameref*{subsubsec:datapipelinestep2} method
works well for single point vibration sources, however is not effective at measuring the vibration information of multiple points with different frequencies, e.g., produced by multiple leaves of a plant. Here, the radar measurement can be modeled as a combination of many independent sources, where the process of separating the mixed signal is introduced in the following section.

\subsubsection{Independent Component Analysis}\label{subsubsec:datapipelinestep3}
As plants contain multiple vibration points at different frequencies (see Section~\ref{subsubsec:vibrationModes}), including adjacent leaves, branches, and the trunk, it becomes necessary to isolate the reflection source that contains the vibration mode of interest. An ideal target leaf vibrates perpendicular to the radar, providing the highest signal-to-noise ratio (SNR) for vibration measurements. However, due to factors like branch architecture and petiole angle, measurements often include many independent vibration sources~\cite{tharwat_independent_2021}. To address this challenge, BSS is a technique used to separate these mixed signals without requiring prior knowledge of the source signals.

Specifically, we formulate the BSS problem as Independent Component Analysis (ICA)~\cite{hyvarinen_independent_2000, tharwat_independent_2021, wu_estimating_2011}, leveraging the virtual antenna pairs commonly available in COTS mmWave radars. For example, our \sn prototype, which uses the AWR1843 radar, consists of 3 transmit (Tx) and 4 receive (Rx) antennas (see Section~\ref{subsubsec:EvaluationMethodology} for details), resulting in $3 \times 4 = 12$ virtual antenna pairs. Here, each antenna pair can be considered a separate virtual vibration sensor since it operates in a different radio channel, and one necessary condition for ICA is that the number of sensors should be equal to or greater than the number of monitored vibration sources.

The ultimate goal of using ICA for BSS in radar signals is to recover a clean, one-dimensional true vibration signal while minimizing other vibrating noises. We achieve this by analyzing data from multiple sensors (i.e., the raw radar data from multiple virtual antenna pairs~\cite{fouad_optimizing_2019}). ICA extracts independent components by maximizing the non-Gaussianity (referred to as "sharpness" or "kurtosis") of the source signal~\cite{tharwat_independent_2021, zhu_complex_2024, wu_estimating_2011, ahmad_review_2011}. By maximizing kurtosis, ICA effectively separates the independent components (leaves, branches, or the trunk), each corresponding to a different source of vibration, from the mixed signals. Evaluating the ICA component with the highest amplitude allows us to identify the leaf with displacement perpendicular to the radar beam, ensuring the highest possible SNR.

Suppose we have a radar with multiple virtual antenna pairs, each of which receives a mixed signal of various plant vibration sources. If we denote the observed radar signal as $\mathbf{x}(t)$, representing a multi-dimensional space corresponding to different virtual antenna pairs, then the received signal can be described as:

\begin{equation}\label{eqn:icamixedsignals}
\mathbf{x}(t) = \mathbf{A} \mathbf{s}(t),
\end{equation}

where $\mathbf{A}$ is the mixing matrix and $\mathbf{s}(t)$ represents the independent source signals.

The goal of the ICA algorithm is to estimate the unmixing matrix $\mathbf{W}$ (such that $\mathbf{W}  \mathbf{A}$ is an identity matrix), after which the estimated source signals $\tilde{\mathbf{s}}(t)$ can be calculated by:

\begin{equation}\label{eqn:icaestimatedsources}
\tilde{\mathbf{s}}(t) = \mathbf{W} \cdot \mathbf{x}(t) = \mathbf{W} \cdot \mathbf{A} \cdot \mathbf{s}(t).
\end{equation}

\sn utilizes the FastICA algorithm~\cite{oja_fastica_2006} to solve Equation~\ref{eqn:icamixedsignals} as it converges much faster than gradient-based descent algorithms~\cite{xu_gait-key_2017,ahmad_review_2011}. FastICA can be used for extracting one independent-component or several independent-components, where the output $\mathbf{w_i} \mathbf{X}$ is decorrelated from previous iterations ($\mathbf{w_1} \mathbf{X}, \mathbf{w_2} \mathbf{X}, \ldots , \mathbf{w_{i-1}} \mathbf{X}$) preventing different output weight vectors from converging to the same optima~\cite{tharwat_independent_2021}.

To extract the estimated source signal using a specific number of independent components, we derive the adjusted signal $\mathbf{s}'(t)$ as follows:

\begin{equation}\label{eqn:icaidealsources}
\mathbf{s}'(t) = \mathbf{W} \bar{\mathbf{s}},
\end{equation}

where $\bar{\mathbf{s}}$ is the matrix of independent components, given by:

\begin{equation}
\bar{\mathbf{s}} = \begin{bmatrix}
\tilde{s}_{11} & \tilde{s}_{12} & \cdots & \tilde{s}_{1N} \\
\tilde{s}_{21} & \tilde{s}_{22} & \cdots & \tilde{s}_{2N} \\
\vdots & \vdots & \ddots & \vdots \\
\tilde{s}_{M1} & \tilde{s}_{M2} & \cdots & \tilde{s}_{MN}
\end{bmatrix},
\end{equation}

where $i = 1, \ldots; M$, $j = 1, \ldots N$; and $M \leq N$. The value of $M$ represents the number of plant vibration sources, and $N$ corresponds to the number of radar samples. Our evaluation in Section~\ref{sec:evaluation} later shows that $M = 12$ in our \sn prototype is sufficient to observe the subtle vibrations produced by different parts (e.g., trunks, branches, and leaves) of the plant.

To find the optimal number of source components ${i}$ using FastICA~\cite{oja_fastica_2006}, we optimize using the absolute value of kurtosis, as shown in Algorithm~\ref{alg:FastICA}. 
Here, Line 1 defines the maximum possible components using the number of virtual antenna pairs  available in the radar. Line 2-4 iterates through the range of components and applies FastICA for each $n$\_components one by one via a loop. Then, Line 5-6 estimates the kurtosis for each independent component in the FastICA model. The number of independent components per model is set by the n\_components parameter in the previous step. 
Line 7-8 calculates the mean absolute kurtosis for each FastICA model. Finally, Line 9 find the optimal number of components where the maximum difference in mean kurtosis occurs, thus maximizing the non-Gaussianity of extracted signals.

\begin{figure}[htbp]
  \centering
  \subfloat[Raw Signals \label{fig:methodICAradarraw}]{\includegraphics[trim={0.3cm 0cm 0.2cm 0.2cm},clip,height=0.25\linewidth]{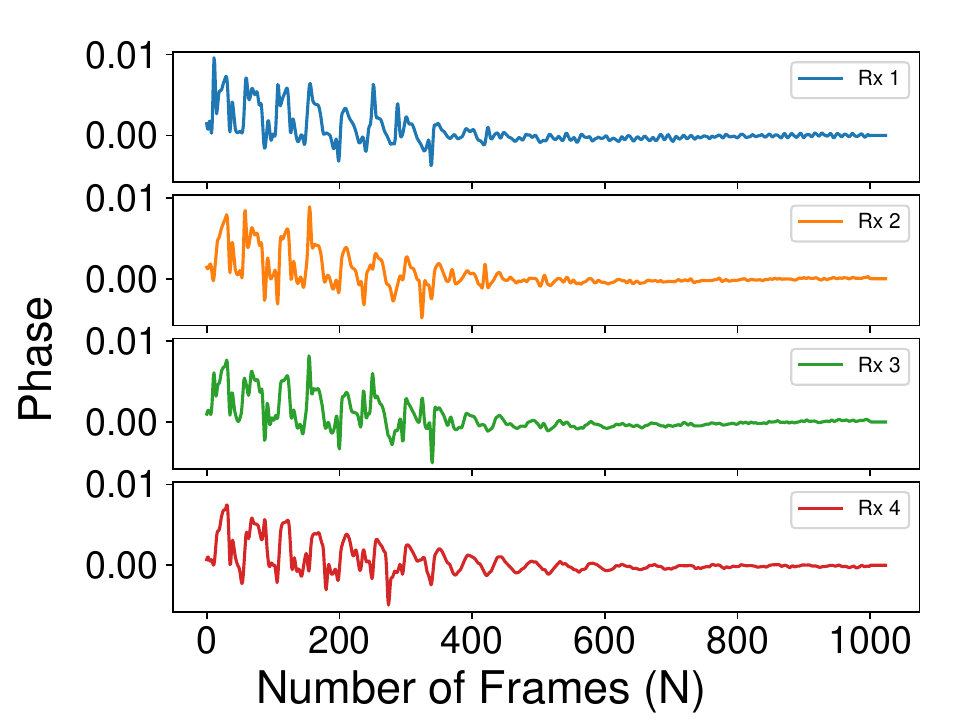}}
  \subfloat[Extracted ICA \label{fig:methodICAradarIC}]{\includegraphics[trim={0.28cm 0cm 0.3cm 0cm},clip,height=0.25\linewidth]{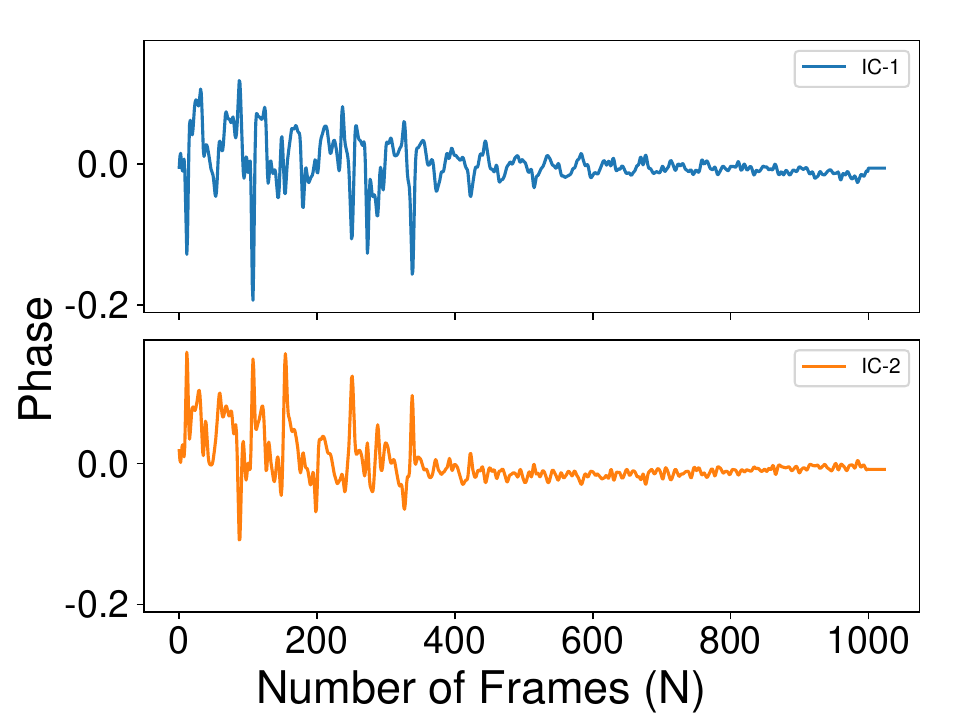}}
  \subfloat[FFT of ICA \label{fig:methodICAfftoutput}]{\includegraphics[trim={0.3cm 0.6cm 0.45cm 0cm},clip,height=0.25\linewidth]{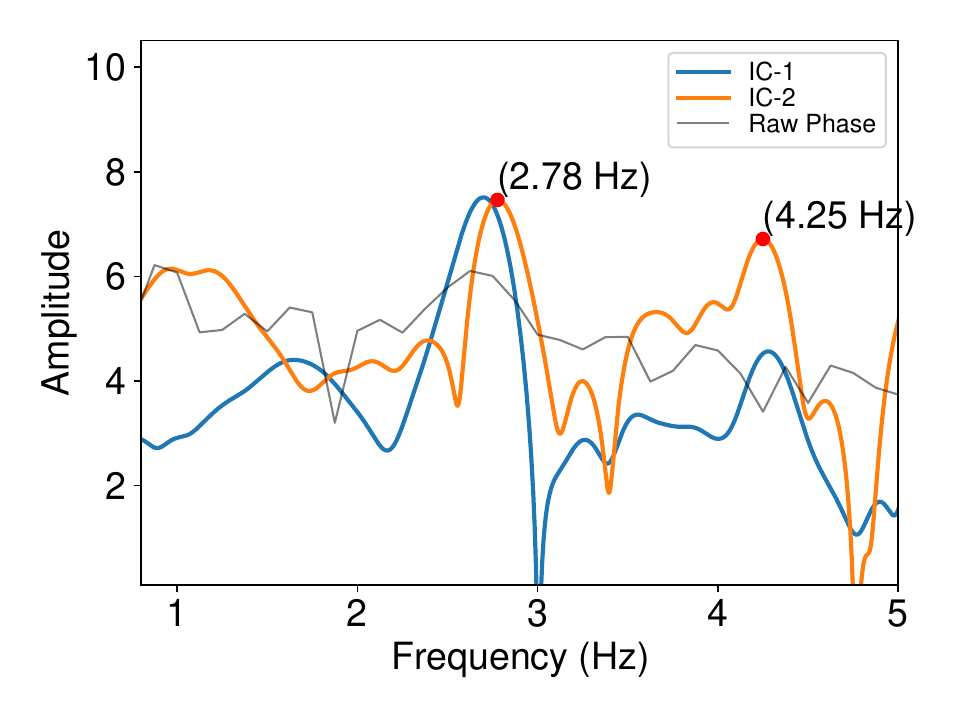}}
  \caption{ICA extract number of source components}
  \label{fig:methodICAresults}
  \Description{}
  \vspace{-1em}
\end{figure}

Figure~\ref{fig:methodICAresults} shows an example that, using the optimal number of independent-components $\bar{\mathbf{s_i}}$, we extract the estimated source signals $\mathbf{s}'(t)$ from Equation~\ref{eqn:icaidealsources}. 
Here, using the vibration measurements after Section~\ref{subsubsec:datapipelinestep2} from different radar virtual antenna pairs as the input to the ICA algorithm (Figure ~\ref{fig:methodICAradarraw}), we can see that the ICA Algorithm~\ref{alg:FastICA} correctly identifies the two vibrating sources at $\approx$ 2.78 Hz and $\approx$ 4.25 Hz respectively and extracts the independent-components, shown in Figure~\ref{fig:methodICAradarIC}. After performing an FFT on each independent-component, the vibration frequency from each leaf can be easily separated, shown in Figure~\ref{fig:methodICAfftoutput}. On the other hand, only one leaf vibration ($\approx$ 2.7 Hz) can be detected if we use the vibration measurements directly without ICA (the blue curve in Figure~\ref{fig:methodICAfftoutput}). 

\begin{algorithm}
\caption{Select Optimal Number of Components using FastICA based on Kurtosis Output}
\label{alg:FastICA}
\KwIn{$x_{ij}$}  \tcp{Input matrix. $x_{ij}$ is $j$-th observation from $i$-th virtual antenna pair}
\KwOut{$C$}      \tcp{Optimal number of components}

$max_{components}$ $\leftarrow$ $x_{i}$\;  \tcp{Maximum possible components is the number of virtual antenna pairs}
$n_{components}$ $\leftarrow$ {1, 2, \ldots, $max_{components}$}\;  \tcp{Range of components to evaluate}

\For{$n \in n_{components}$}{
    $icaModels[n] \leftarrow$ \text{FastICA(n\_components=n).fit\_transform($x_{ij}$)}\;  \tcp{Fit FastICA model for each $n$ and store the transformed data}
}

\For{$model \in icaModels$}{
    $kurtosisValues[model] \leftarrow$ \text{kurtosis(model, axis=0, fisher=True)}\;  \tcp{Calculate kurtosis for each component in the model}
}

\For{$kurt \in kurtosisValues$}{
    $kurtosisMeans[kurt] \leftarrow$ \text{mean(abs(kurt))}\;  \tcp{Compute the mean absolute kurtosis for each model}
}

$C$ $\leftarrow$ \text{argmax(diff(kurtosisMeans))} + 1\;  \tcp{Find the optimal number of components using the maximum difference in mean kurtosis}

\Return $C$\;

\end{algorithm}

\section{Evaluation}\label{sec:evaluation}


\subsection{Goals, Methodology and Metrics}\label{subsec:goalsMethodologyMetrics}

\subsubsection{Goals}
Our evaluation has three primary goals: first, assess whether \sn can effectively monitor plant water stress; second, evaluate the accuracy of \sn in estimating leaf vibration frequencies by benchmarking it against the state-of-the-art approach~\cite{jiang_mmvib_2020}; and third, understand the contributions of different components in \sn's signal processing pipeline.

\subsubsection{Methodology}
\label{subsubsec:EvaluationMethodology}
\paragraph{Prototype}

To evaluate the performance of \sn, we developed a prototype using a COTS AWR1843Boost radar~\cite{texas_instruments_awr1843boost} and a DCA1000EVM data collection board~\cite{texas_instruments_dca1000evm}, both from Texas Instruments (TI). The AWR1843 is a 77 GHz mmWave radar featuring 3 transmit antennas (Tx1, Tx2, and Tx3) and 4 receive antennas (Rx1, Rx2, Rx3, and Rx4). The DCA1000EVM enables real-time data capture from the AWR1843 to a mini-PC~\cite{beelink_sei12}, storing the raw ADC values in a binary file. Table~\ref{tab:radarchirpparameters} details the FMCW chirp parameters utilized in the \sn prototype.
\begin{table}
\centering
  \caption{TI AWR1843 Radar Chirp Parameters}
  \label{tab:radarchirpparameters}
\begin{tabular}{|l||l|} 
\toprule
Parameters & Specifications \\ 
\hline
Frequency Range (GHz) & 77-81 \\
Frequency slope ($MHz/\mu s$) & 99.987 \\
Idle time ($\mu s$) & 7 \\
ADC sampling frequency (ksps) & 10,000 \\
ADC start time ($\mu s$) & 7 \\
Number of chirps per profile & 128 \\
Ramp end time ($\mu s$) & 40 \\
Effective chirp time ($\mu s$) & 47 \\
Number of ADC samples & 256 \\
Frame length (ms) & 8 \\
Bandwidth (GHz) & 2.56 \\
Range resolution (cm) & 5.86 \\
\bottomrule
\end{tabular}
\end{table}

\paragraph{Leaf types}

We use four Avocado plants (Persea americana) for both our in-lan and on protected cropping farm experiments. 
The plants were grown in pots 10-inch wide and 10-inch deep, using a generic off-the-shelf soil potting mix with no additional fertilizer.

\paragraph{Experimental Setup}
Unless specified otherwise, the AWR1843 radar, DCA1000 board, and mini-PC were placed in a plastic enclosure suspended from a metal frame above the plant, with distance to the target leaf approximately 0.5 m.  A linear solenoid~\cite{core_electronics_solenoid_36v} was placed near the base of the main trunk to provide a controlled vibration force which can be set by using a voltage. 


\subsubsection{Metrics}
We use the MAE in Hertz (Hz) to evaluate the performance of \sn, which is calculated as the difference between the estimated and actual vibration frequencies.

\subsection{In-lab Experiments}
\subsubsection{Performance comparison with SOTA}
We conducted in-lab experiments to study the vibration frequency of the Avocado plants. Here, we collected the mmWave radar $I/Q$ measurements from our \sn prototype, together with the ground truth produced by in lab high precision lasers,  
during induced damped plant vibration twice per day (i.e., day and night) to observe their diurnal cycles (see Section~\ref{sec:diurnalcycle}).
Furthermore, we bench-marked the performance of \sn against that of mmVib~\cite{jiang_mmvib_2020} and naive raw \textbf{Radar} displacement calculation method, i.e., Eq.~(\ref{eq:d_n}). 

Table \ref{tab:mae_results} highlights the performance of \sn for different frequency ranges. For the 1-2 Hz range, the MAE is 0.0119 Hz, indicating high precision and low error in this frequency range. In the 2-3 Hz range, the MAE increases to 0.0219 Hz, showing a moderate increase in error but acceptable for leaf frequency monitoring. The 3-4 Hz range has a significantly higher MAE of 0.0651 Hz, showing a noticeable increase in error and variability. For the 4-5 Hz range, the MAE is 0.0297 Hz, indicating a reduction in error compared to the 3-4 Hz range but still higher than the lower frequency ranges. Finally, the overall MAE for the combined 0-5 Hz range is 0.0176 Hz, which is relatively low when considering the entire frequency range, though it varies within specific sub-ranges.

\begin{figure}[htbp]
  \centering
 \subfloat[Multi-leaf performance\label{fig:resultsavocadomultipleleafica}]{
 \includegraphics[height=0.37\linewidth]{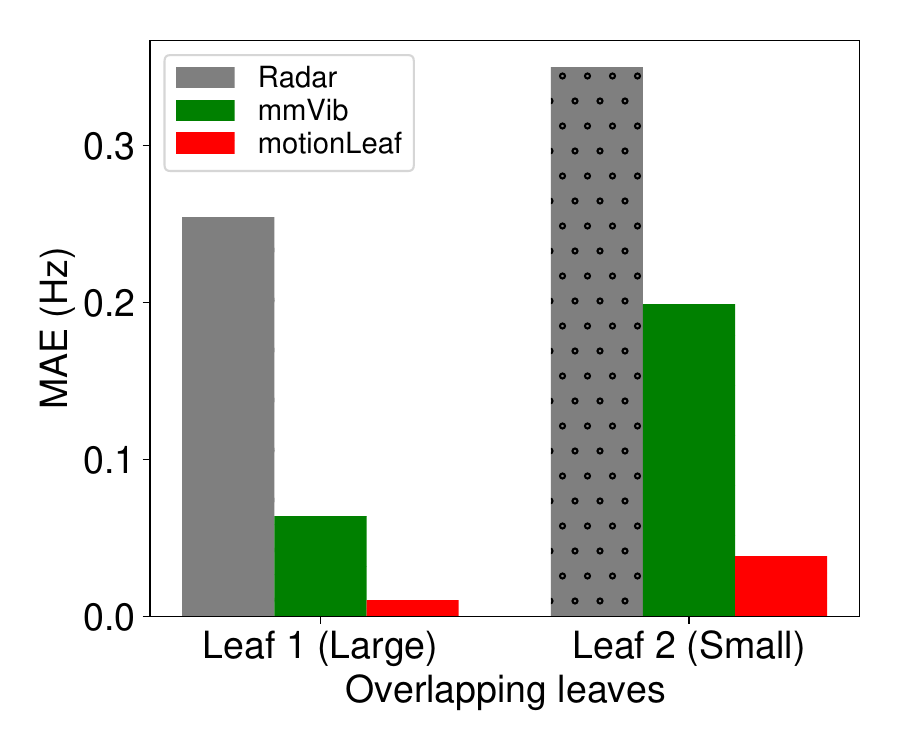}}
 \hspace*{1em}
 \subfloat[Module Performance 
  \label{fig:resultsablationmoduleblocksmae}]{\includegraphics[height=0.37\linewidth]{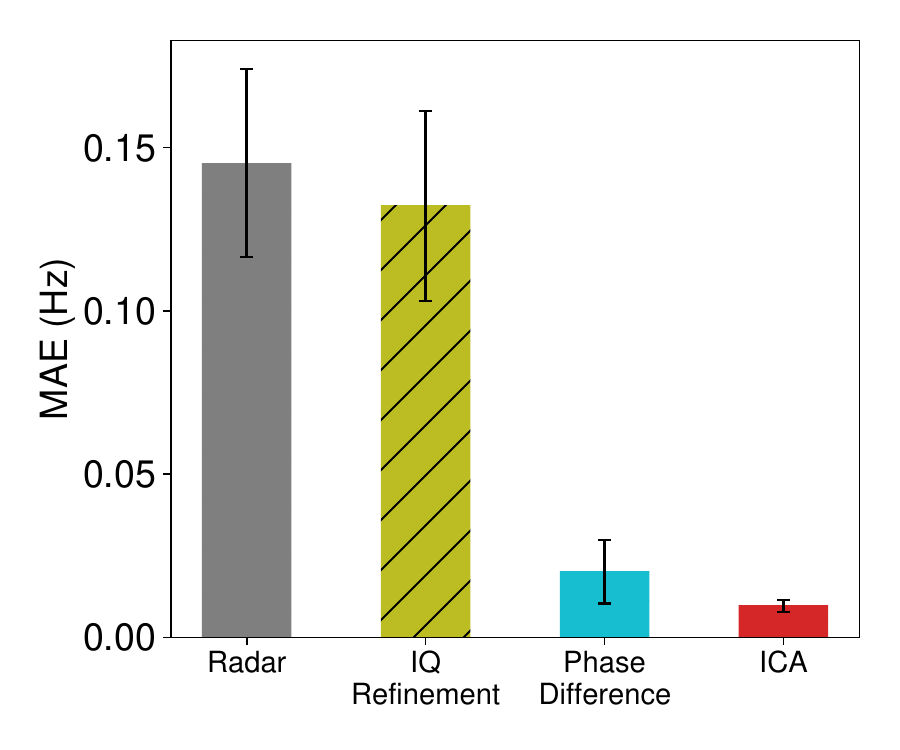}}
  \caption{Evaluation of \sn performance}
  \label{fig:resultsmultipleleaficaevaluation}
  \Description{}
  \vspace{-1em}
\end{figure}

The table also show that \sn outperforms mmVib across all frequency ranges. We see that the accuracy of the vibration measurement decreases as the frequency increases, highlighting the challenges in maintaining precision at higher frequencies. This is mostly due to the lower signal strength coming from smaller leaves, which vibrate at higher frequencies, compared to larger leaves with vibrate at lower frequencies. For example, Figure~\ref{fig:resultsavocadomultipleleafica} shows the results from a sample vibration measurement with two overlapping leaves (Leaf 1, a larger leaf, and Leaf 2, a smaller leaf). For both leaves, the raw radar data shows the highest MAE, approximately 0.25 Hz for Leaf 1 and 0.35 Hz for Leaf 2, indicating substantial measurement errors without advanced processing techniques. The mmVib method significantly reduces the error, bringing it down to around 0.06 Hz for Leaf 1 and 0.19 Hz for Leaf 2, demonstrating an improved accuracy in detecting leaf vibrations. However, the \sn system outperforms both approaches, achieving the lowest MAE at approximately 0.01 Hz for Leaf 1 and 0.03 Hz for Leaf 2. These results illustrate the superior performance of the \sn system in accurately isolating and measuring the vibrations of individual overlapping leaves, confirming its efficacy in complex scenarios where multiple vibration sources are present.

\begin{table}[h!]
\centering
\caption{Comparison of MotionLeaf with SOTA}
\begin{tabular}{|c|c|c|c|}
\hline
Frequency & Radar & mmVib & \sn \\ \hline
1-2 Hz & 0.1405 $\pm$ 0.014 & 0.0242 $\pm$ 0.005 & 0.0119 $\pm$ 0.002 \\ \hline
2-3 Hz & 0.2047 $\pm$ 0.034 & 0.0257 $\pm$ 0.007 & 0.0219 $\pm$ 0.004 \\ \hline
3-4 Hz & 0.1432 $\pm$ 0.068 & 0.2139 $\pm$ 0.066 & 0.0650 $\pm$ 0.012 \\ \hline
4-5 Hz & 0.1966 $\pm$ 0.107 & 0.1118 $\pm$ 0.063 & 0.0297 $\pm$ 0.006 \\ \hline
All & 0.1487 $\pm$ 0.013 & 0.0416 $\pm$ 0.009 & \textbf{0.0176 $\pm$ 0.002} \\ \hline
\end{tabular}
\label{tab:mae_results}
\end{table}

\subsubsection{Ablation study}
To understand the contributions of each component in the signal processing pipeline of \sn, we conducted an ablation study. The components evaluated include Signal Phase Refinement, Coherent Phase Difference, and ICA. Figure~\ref{fig:resultsablationmoduleblocksmae}
shows the MAE for different components of the \textit{MotionLeaf} system. The raw radar data shows the highest MAE at approximately 0.14 Hz, indicating a baseline error when no additional processing is applied. The $I/Q$ Refinement module slightly reduces the error to about 0.12 Hz, demonstrating a modest improvement in measurement accuracy by refining the $I/Q$ components. The Phase Difference method significantly decreases the MAE to around ~0.02 Hz, highlighting the effectiveness of phase difference calculations in capturing more accurate vibration frequencies. Finally, the ICA module achieves the lowest MAE at approximately ~0.01 Hz, underscoring its superior capability in isolating and identifying the independent vibration sources, thus providing the most accurate measurement. These results illustrate the cumulative benefit of each processing step in the \textit{MotionLeaf} system, with the most substantial improvements attributed to the Phase Difference and ICA techniques.

\subsubsection{Impact of the Radar-Leaf Distance}
Figure~\ref{fig:resultsavocadosingleleafdistance} shows the MAE in Hz of \sn at varying distances (50 cm, 80 cm, and 110 cm) for measuring leaf vibration frequencies. \sn exhibited a very low MAE (approximately 0.01 Hz) at 50 cm, which increased (approximately 0.22 Hz) at 80 cm and the trend continued at 110 cm with the highest MAE (approximately 0.32 Hz). The results indicate that \sn is very accurate under 50cm, with frequency errors acceptable for measuring frequency changes in water-stressed leaves. This highlights the suitability of using \sn, particularly within certain distances, for mmWave radar remote plant health monitoring.

\subsubsection{Impact of Vibration Force}
\begin{figure}[htbp]
  \centering
  \subfloat[Impact of distance\label{fig:resultsavocadosingleleafdistance}]{\includegraphics[height=0.39\linewidth]{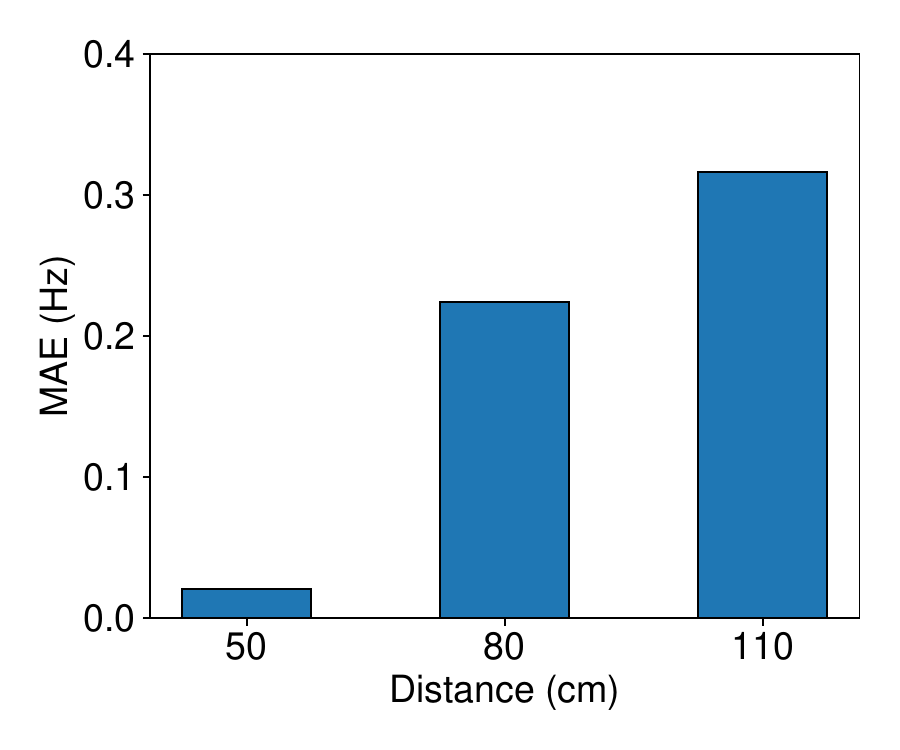}}
  \hspace*{1em}
  \subfloat[Impact of plant size
  \label{fig:resultsavocadosingleleamultihammer}]{\includegraphics[height=0.39\linewidth]{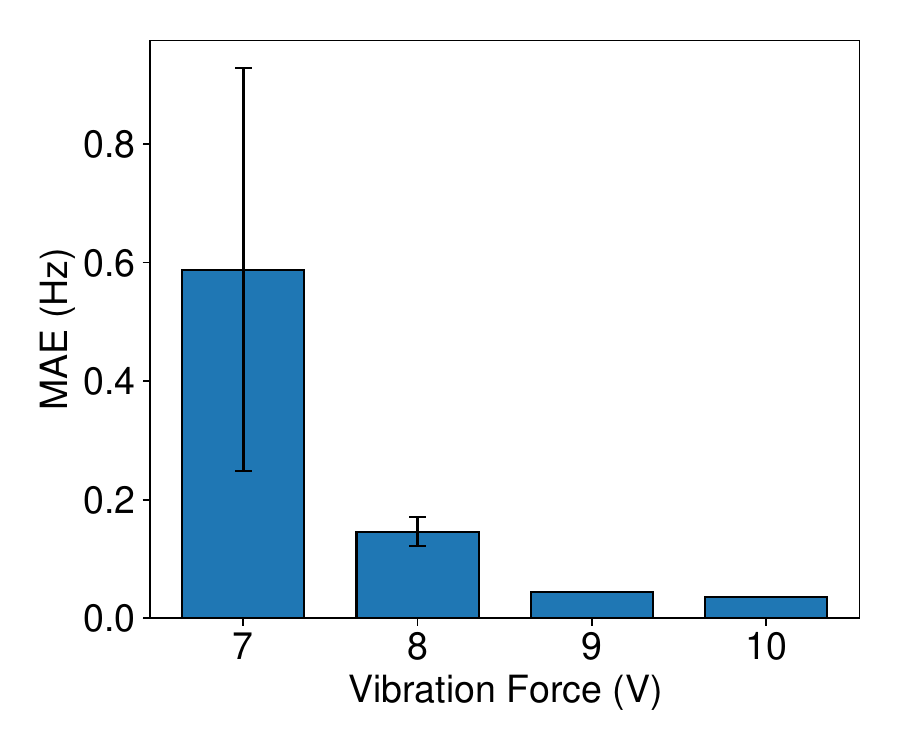}}
  \caption{Evaluation of leaf distance and vibration input power}
  \label{fig:resultsleafdistanceandvibrationevaluation}
  \Description{}
  \vspace{-1em}
\end{figure}

Figure~\ref{fig:resultsavocadosingleleamultihammer} shows the MAE in Hz of frequency measurements for four different Avocado plants using varying solenoid voltages (7 V, 8 V, 9 V, and 10 V). The results show a high MAE at 7 V (approximately 0.60 Hz), indicating that force applied is not providing enough movement in the leaves to enable accurate displacement measurements. At 8 V we see a substantial drop in MAE (approximately 0.145 Hz), however this is still above the acceptable error tolerance where the diurnal frequency range requires less than 0.1 Hz (see Figure~\ref{fig:resultsavocadodeltadiurnalfrequency} later for the details). Using higher forces of 9V and 10V exhibited significantly lower errors, (approximately 0.02 Hz and approximately 0.01 Hz respectively), suggesting that there is a minimum vibration force required to achieve accurate measurements. These results imply that the accuracy of leaf vibration frequency measurements is dependent on the applied vibration force. For optimal accuracy and to minimize error, forces above 10 V should be applied.

\subsubsection{Impact of Chirp Repetition Time on Phase Difference Module}

The impact of chirp repetition time on IQR filtered data shows a clear trend of increasing data variability with longer chirp repetition times. At shorter times (50 µs to 200 µs), the IQR filtered data remains below 0.02 radians, indicating minimal noise and high signal quality. However, as the chirp repetition time increases to 800 µs, 1,600 µs, and especially 3,200 µs, the IQR filtered data rises sharply, reaching ~0.28 radians at 3,200 µs. This suggests that the damped vibration of a plant is subtle and longer chirp repetition times introduce uncertainty in the phase difference measurements between chirps, degrading the signal quality.

The corresponding impact on MAE shows a similar trend. At shorter chirp repetition times (i.e., 50 µs to 400 µs), the MAE remains consistently low, around 0.01 Hz to 0.015 Hz, indicating high accuracy in frequency estimation. As the chirp repetition time increases beyond 800 µs, the MAE begins to rise, reaching approximately 0.025 Hz at 3,200 µs. This correlation between increased chirp repetition time, higher IQR filtered data variability, and rising MAE highlights the importance of optimizing chirp repetition time to capture vibrating objects. Shorter chirp repetition times are preferable for maintaining low MAE and high-quality phase difference measurements, crucial for accurate plant vibration monitoring.
\begin{figure}[htbp]
  \centering
  \subfloat[IQR Filtering algorithm\label{fig:resultsablationmodulechirpsiqr}]{\includegraphics[height=0.39\linewidth]{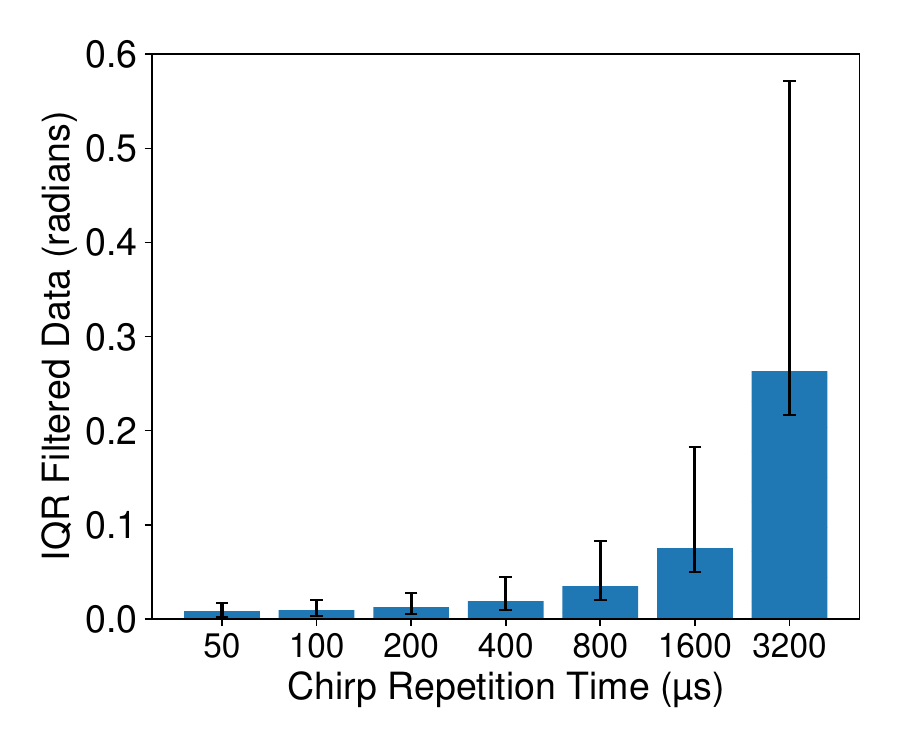}}
  \hspace*{1em}
  \subfloat[System performance 
  \label{fig:resultsablationmodulechirpsmae}]{\includegraphics[height=0.39\linewidth]{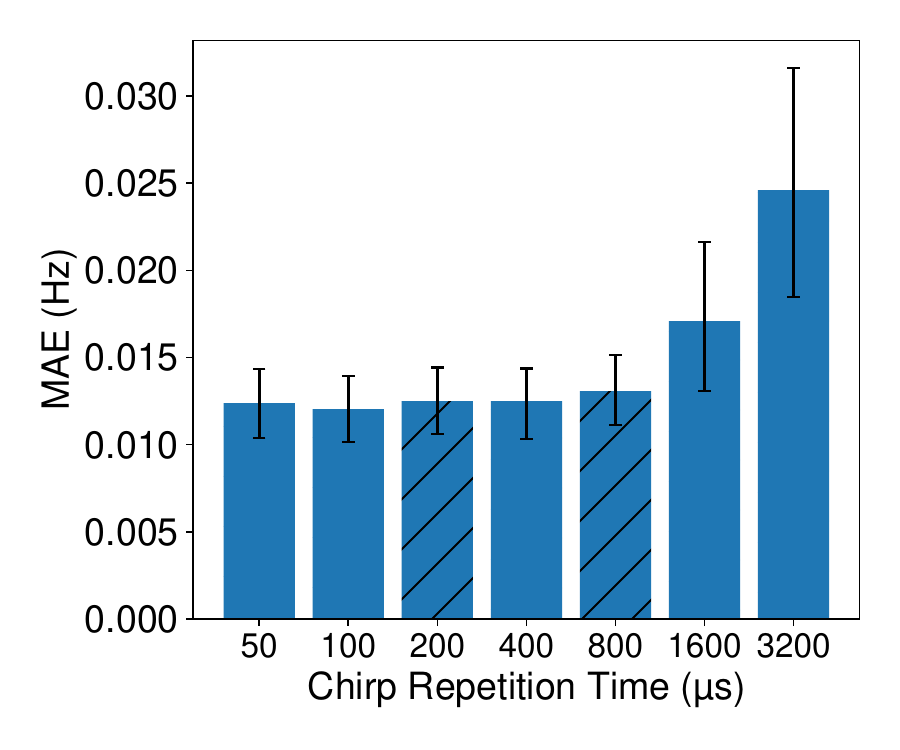}}
  \caption{Impact of different Chirp Repetition Times}
  \label{fig:resultsablationmodulechirps}
  \Description{}
\end{figure}

\subsection{Protected Cropping Farm Experiments}
To study the plant water stress monitoring performance of \sn, we conducted a drought experiment to study the vibration frequency differences of a plant between day and night (i.e., diurnal cycles). A larger frequency difference indicates that the plant has a better ability to recover from water stress and maintain its physiological functions, as discussed in Section~\ref{sec:diurnalcycle} earlier.

\begin{figure}[htbp]
  \centering
  \includegraphics[trim={0cm 0cm 0cm 3cm},clip,height=0.5\linewidth]{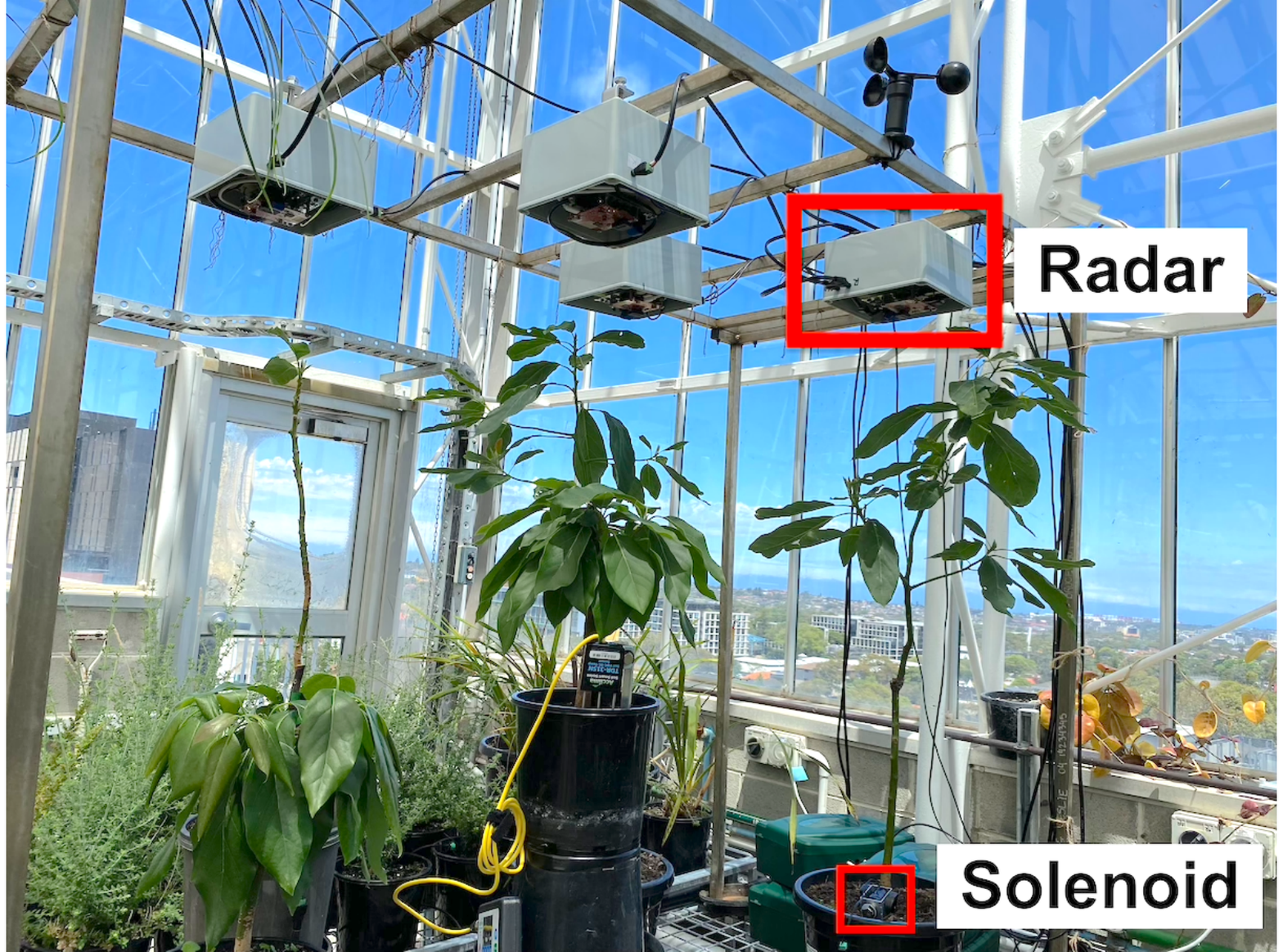}
  \caption{Farm experimental setup}
  \label{fig:methodavocadooutdoorplant}
  \Description{}
  \vspace{-1em}
\end{figure}

\paragraph{Data Collection}
We used four Avocado plants, initially watering each plant and then measuring them for 7 days without additional hydration (see Figure~\ref{fig:methodavocadooutdoorplant}). An automated script ran on the mini-PC, taking radar measurements every hour synchronized with solenoid actuation. Additionally, we recorded temperature, humidity, and wind speed using COTS Internet of Things (IoT) sensors~\cite{arduino_nano_33_ble, dfrobot_wind_speed_sensor} every 15 seconds. Our results indicate that the temperature was ($18^{\circ}C \pm 5.4^{\circ}C$), humidity was ($69\% \pm 13\%$), and wind speed was ($0.1\,m/s \pm 0.21\,m/s$) during our experiment.

\begin{figure}[htbp]
  \centering
    \subfloat[Day/Night Hz delta\label{fig:resultsavocadodeltadiurnalfrequency}]{\includegraphics[trim={0.3cm 0.6cm 0.2cm 0cm},clip,height=0.37\linewidth]{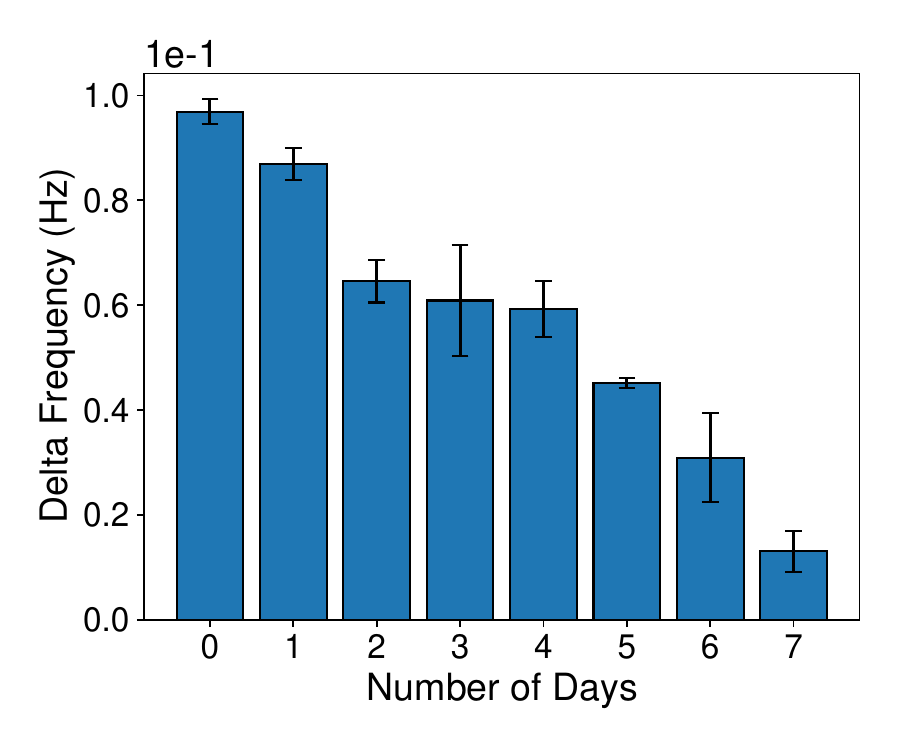}}
  \hspace*{0.1em}
  \subfloat[Natural frequency
  \label{fig:resultsvibration3freqhammwindow}]{\includegraphics[trim={0cm 0cm 0.2cm 1.8cm},clip,height=0.37\linewidth]{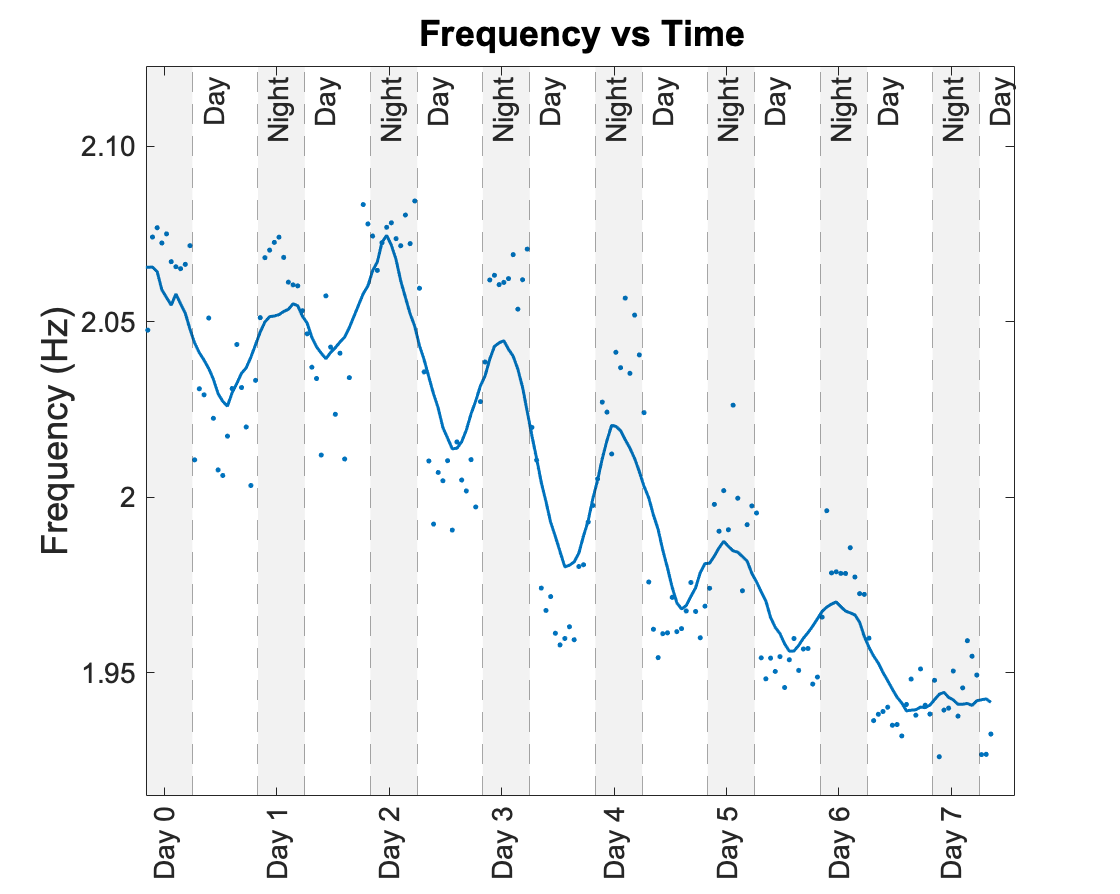}}
  \caption{Drought experiment results for Avocado plant}
  \label{fig:evaluationsetup}
  \Description{}
  \vspace{-1em}
\end{figure}

Figure~\ref{fig:resultsavocadodeltadiurnalfrequency} illustrates the variation in day/night frequency of one of the Avocado plants during a 7-day drought experiment. The y-axis represents the delta frequency in Hz, calculated as the median frequency during the day minus the median frequency during the night. The x-axis indicates the number of days since the plant was last watered. Initially, on Day 0, the delta frequency is at its highest, close to 0.1 Hz. As the days progress without watering, there is a noticeable decline in the diurnal frequency delta. By Day 3, the delta frequency has decreased to approximately 0.06 Hz, indicating a significant reduction. This trend continues, with a further drop observed on subsequent days, reaching around 0.012 Hz by Day 7. The consistent downward trend in delta frequency suggests that the plant's turgor pressure and overall health are declining due to the lack of water, validating the effectiveness of \sn in detecting water stress through changes in leaf vibration frequencies.

Figure~\ref{fig:resultsvibration3freqhammwindow} displays the raw natural vibration frequency values of one plant used to calculate the delta frequency discussed earlier, produced by \sn. The figure shows that the frequency has a downward trend under water stress, similar to  Figure~\ref{fig:methodspringconstantleaftable}. Results from the other three Avocado plants exhibit a similar trend and are omitted for brevity.

\section{Related Work}\label{sec:relatedwork}

\sn is related to a wide-range of research areas and we group the related works into two key areas consisting of: (i) \nameref*{itm:rw_1} and (ii) \nameref*{itm:rw_2}.

\subsection{Vibration measurements}\label{itm:rw_1}

Recent papers have explored using mmWave radar for vibration monitoring. For example, Multi-Vib~\cite{yang_multi-vib_2022} employs a single mmWave radar to sense the vibration displacement of multiple points via signal reflection. mmEavesdropper~\cite{feng_mmeavesdropper_2023} utilizes signal augmentation techniques, like Chirp-Z transform, to enhance the accuracy of the micro-vibration signal. mmVib~\cite{jiang_mmvib_2020} segmented the chirp to exploit different center frequencies. Dancing Waltz with Ghosts~\cite{guo_dancing_2021} explored the multipath reflection principle of RF signals to receive the line-of-sight (LOS) and non-line-of-sight (NLOS) vibrations from rotating machinery to detect 2D vibration movements. However, this method relies on a large flat reflecting surface (e.g., wall and floor) close to the rotating equipment, making it overly restrictive for plant vibration monitoring.

\subsection{Plant dynamics}\label{itm:rw_2}
The study of plant dynamics, particularly in response to environmental stressors, has been an area of significant interest~\cite{de_langre_plant_2019}. Various works have looked at modeling plant dynamics to infer physiological properties, with vibration analysis already been used as part of the agriculture processes~\cite{de_langre_plant_2019, afzal_leaf_2017}. 

Exploring the mechanical properties of the plant, several studies \cite{caliaro_effect_2013, gonzalez-rodriguez_turgidity-dependent_2016, kanahama_rigidity_2023} have analyzed the bending stiffness of leaves under drought stress, revealing that cell turgor loss contributed to changes in flexural rigidity during water stress conditions. Recent works have specifically explored the frequency patterns of plants under stress, proposing that changes in vibration frequency could serve as indicators of the plant's water status~\cite{khait_sounds_2023, caicedo-lopez_effects_2020}.

\section{Conclusion and future work}\label{sec:conclusion}

In this paper, we introduced \sn, an innovative multi-point vibration measurement system utilizing low-cost mmWave radar for monitoring plant water stress. Our study demonstrates that leaf vibration frequency can serve as a reliable, non-destructive indicator of water stress in agricultural crops. Through the analysis of diurnal vibration patterns of Avocado plants under drought conditions, we validated the efficacy of this approach in detecting plant health status. While our results are promising, further research is required to assess the generality of this method to a variety of plant species. The non-invasive and real-time monitoring capabilities of \sn suggest significant potential for future agricultural systems, offering a novel tool for precision agriculture and plant health management.

\begin{acks}
    This work is supported in part by the Future Food Systems under Grant Fund RE932. 
    This research includes computations using the computational cluster Katana supported by Research Technology Services at UNSW Sydney.

    Statement: During the preparation of this work the author used OpenAI's GPT-4o to improve the readability of various sections. After using this tool/service, the author reviewed and edited the content as needed and take full responsibility for the content of the publication.

\end{acks}

\bibliographystyle{ACM-Reference-Format}
\bibliography{bibliography/ch4-vibrations, bibliography/websites}
\end{document}